\documentclass[%
 reprint,
 amsmath,amssymb,
 aps,
]{revtex4-2}

\usepackage{makecell}
\usepackage{graphicx}
\usepackage{dcolumn}
\usepackage{bm}
\usepackage[normalem]{ulem}
\usepackage{hyperref}
\usepackage{natbib}
\vspace{2cm}

\begin{document}

\title{Enabling Technologies for Scalable Superconducting Quantum Computing}
\author{Xanthe Croot}
\thanks{These authors contributed equally to this work.}
\affiliation{University of Sydney}

\author{Kasra Nowrouzi}
\thanks{These authors contributed equally to this work.}
\affiliation{Lawrence Berkeley National Laboratory}

\author{Carmen G. Almudever}
\affiliation{Universitat Politècnica de València}

\author{Alexandre Blais}
\affiliation{Université de Sherbrooke and CIFAR}

\author{Malcolm Carroll}
\affiliation{IBM Quantum}

\author{Edoardo Charbon}
\affiliation{École Polytechnique Fédérale de Lausanne}

\author{Jerry Chow}
\affiliation{IBM Quantum}

\author{Vivek Chidambaram}
\affiliation{National Quantum Computing Centre}

\author{Andrew N. Cleland}
\affiliation{University of Chicago}

\author{David Danovitch}
\affiliation{Université de Sherbrooke}

\author{Joseph Emerson}
\affiliation{Keysight}

\author{Daniel Friedman}
\affiliation{IBM Quantum}

\author{Masao Tokunari}
\affiliation{IBM Quantum}


\author{David Gunnarsson}
\affiliation{Bluefors}

\author{Raymond Laflamme}
\affiliation{University of Waterloo}

\author{John Martinis}
\affiliation{University of California, Santa Barbara}

\author{Robert McDermott}
\affiliation{University of Wisconsin-Madison}

\author{William D. Oliver}
\affiliation{Massachusetts Institute of Technology}

\author{Michel Pioro-Ladrière}
\affiliation{Nord Quantique}

\author{Yoshiaki Sato}
\affiliation{TDK Corporation}

\author{Hidenori Ohata}
\affiliation{TDK Corporation}

\author{Kouichi Semba}
\affiliation{The University of Tokyo}

\author{Christopher Spitzer}
\affiliation{Lawrence Berkeley National Laboratory}

\author{Irfan Siddiqi}
 \email{irfan\_siddiqi@berkeley.edu}
\affiliation{University of California, Berkeley}


\date{\today}

\begin{abstract}
Experiments with superconducting quantum processors have successfully demonstrated the basic functions needed for quantum computation and evidence of utility, albeit without a sizable array of error-corrected qubits. The realization of the full potential of quantum computing centers on achieving large scale fault-tolerant quantum computers. Science, engineering and industry advances are needed to robustly generate, sustain, and efficiently manipulate an exponentially large computational (Hilbert) space as well as supply the number and quality components needed for such a scaled system. In this article, we suggest critical areas of quantum system and ecosystem development, with respect to the handling and transmission of quantum information within and out of a cryogenic environment, that would accelerate the development of quantum computers based on superconducting circuits.  

\end{abstract}

\maketitle



\section{\label{sec:level1} Introduction}

Nearly a century after the mathematical and philosophical foundations of quantum physics were laid, we are now embarking on the challenge of producing technologies that leverage uniquely quantum phenomena to execute information processing tasks that are beyond the reach of conventional classical hardware. The basic functional elements needed for universal quantum computation have been demonstrated in proof-of-concept experiments. Evidence of utility is also emerging. 

More qubits with a higher degree of coherence than currently available is a common goal to move towards the full promise of large scale fault-tolerant quantum computation (FTQC). Significantly larger machines are anticipated than in quantum computers that exist today. Supporting the pursuit of improved qubits in a scaled environment, advances and ecosystem development is sought for an entire suite of technologies that provide the capability to efficiently shuttle signals and information on and off a quantum processor. 


\begin{figure*}[] 
    \centering
    \includegraphics[width=0.9\linewidth, clip,trim= 0mm 0mm 0mm 0mm]{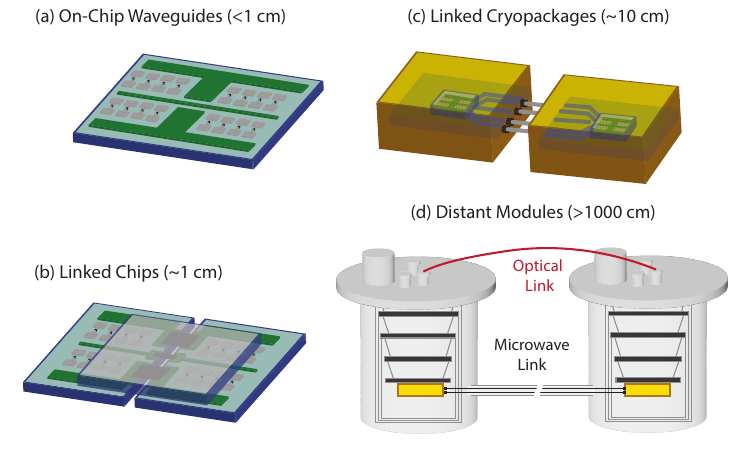}
    \caption{Modular QPU Approaches: (a) Interconnected processing units linked by on-chip microwave transmission lines, (b) separate QPU chips connected by a linker chip, (c) independently packaged QPUs connected by cabling in an single cryostat, and (d) independent QPUs linked by either native microwave frequency links of approximately one or more meters, or via conversion to itinerant optical photons in different cryostats for greater than 1000 cm.}
    \label{fig:modularity}
\end{figure*}

In this article, we focus on superconducting qubits and consider technologies that would accelerate scaling to a significantly larger number of qubits for future large scale fault-tolerant quantum error corrected systems greater than 100-1000 physical qubits. In this discussion we focus on the interface with a QPU rather than on the  materials science and quantum electrical engineering needed to produce an array of highly-coherent, highly-entangled qubits. We first examine some motivations for modular architectures; we then explore the cryogenic back-plane where a QPU is placed; the need for cryogenic electronics; the classical controls needed to operate a quantum computer; and finally the classical computing resources that need to be paired with a quantum computer. This report is the outcome of discussions that began among participants of the Roadmap Workshop on Developing the Complementary Technologies to Enable Quantum Computing, hosted by the CIFAR Quantum Information Science program.

\section{\label{sec:level1} Modular Quantum Processors} 



\subsection{Defining a Module}

Increasing the number of superconducting qubits on a single substrate die heightens challenges related to qubit yield (including performance uniformity), qubit frequency collisions, and wiring complexity. There are, therefore, advantages to considering modular qubit architectures, which can range from attaching multiple qubit dies (``chiplets") to an interconnect substrate providing control and readout wiring as well as inter-die connectivity, to placing qubit dies in connectorized packages and using cabling to provide the needed inter-die connections. Similar modular designs can be used for the control and readout stacks between the qubit modules and the room-temperature end of the cryostat, simplifying the complexity of each module-specific stack, while retaining quantum-coherent communication linkages between the qubit modules. If these linkages are maintained at sufficiently low cryogenic temperatures, these can use microwave signals to achieve and maintain quantum coherent connectivity \citep{magnard_microwave_2020}; this approach, however, is not conceptually distinct from simply building a larger cryostat to accommodate larger circuit designs. If, instead, room temperature linkages are needed, for instance to link modules in  distant cryostats, quantum coherent transduction between microwave and optical frequencies would be required.

\subsection{Engineering/Implementation Considerations}

Different versions of these levels of modularity have been explored, including dies mounted on a substrate \citep{gold_entanglement_2021}, qubits connected by on-chip waveguides \citep{zhong_deterministic_2021}, packaged qubit circuits connected to one another with moderate-length cabling \citep{zhong_deterministic_2021}, qubits linked by a $64$ m long cable within a conventional cryostat \citep{QIU2024}, to qubits with completely separate control, readout and cryogenic stacks linked by a 30 m cryogenic link used to demonstrate a loophole-free Bell inequality violation \citep{storz_loophole-free_2023}. We note that in this context, different link lengths lead to different communication modalities: Very short links, with lengths less than a few wavelengths at the qubit communication frequency allow direct qubit-qubit swaps, while for longer cable links, swaps via resonant cable modes or shaped envelopes for itinerant (mobile) photon signals are preferred. While swaps using ``dark modes'' can sidestep losses associated with low quality-factor cabling \citep{leung_deterministic_2019}, these tend to be slower than direct ``bright mode'' swaps or itinerant signals. ``Bright mode" swaps or itinerant signals, however, require very high quality factor cables and cable-module connections \citep{zhong_deterministic_2021,niu_low-loss_2023}, which present challenges for longer-range direct coupling.

An important future consideration for longer cable links will also be the development of `hot swap' cable-connect configurations (e.g. at 4K or mK) that would allow isolated fridge modules to be brought down for maintenance without stopping the entire system. 

For longer connections outside a cryostat, optical fibers or free-space transmission of optical signals are highly appealing approaches. However, using these for quantum-coherent communication will require a means of transducing between microwave signals generated by superconducting qubits and optical signals for transmission, for which there are a number of approaches being pursued, mostly based on methods of generating microwave-frequency sidebands on an optical carrier using optomechanical or electrooptic methods \citep{meesala_quantum_2024, kumar_quantum-enabled_2023, fan_superconducting_2018, xu_bidirectional_2021, sahu_quantum-enabled_2022, higginbotham_harnessing_2018, jiang_optically_2023, zhao_electro-optic_2023}.
How such communication channels will be used will depend on the available fidelities and data rates, which will also determine whether these links can serve to extend a quantum computation over large distances, such as using teleported gates \citep{ang_arquin_2024}, or if instead these are a means to securely communicate private information \citep{Gisin2007}. We note that most optical communication schemes rely on good long-term memories at the endpoints \citep{Gisin2007}, for instance provided by sufficiently error-protected logical qubits.





\subsection{Quantum Error Correction and Modularity}




While qubit modularity offers relaxed requirements for qubit uniformity and module wiring, it can significantly complicate the picture when considering quantum error correction. This leads to research questions of optimal distribution of logical capability and communication. Multiple modules, with possibly separate control and readout stacks might, for example, provide a backbone of quantum memory \citep{fowler_surface_2012, bravyi_high-threshold_2023, Michael2016}, whose size scales with the number of modules, while other logical functionality might be distributed to its own modules or integrated in some fashion with the memory modules. Regardless of the architecture choices, arranging for fault-tolerant operations across modules introduces challenges and a complex trade space. The module-to-module interconnects ideally need to extend the warp and weft of the code fabric used within each module, in line with and uninterrupted by the module boundaries. The inter-module interconnect densities and fidelities would also need to be sufficient to maintain the effective code distance and functionality introducing performance and yield challenges for hardware and architectural choices. Furthermore, control and readout electronics, error decoding software, and the algorithmic compilation software, would all have to be designed and operated to account for these module boundaries. 

One way to reduce the challenge is to reduce the number of modules and interfaces through reducing the number of qubits per logical qubit. Such a direction also reduces qubit yield requirements. Recent work on high code rate quantum low-density parity-check (qLDPC) codes has recently identified more hardware implementable choices \citep{bravyi_high-threshold_2023, xu2023constantoverheadfaulttolerantquantumcomputation}. Complementing other architecture works relying on more orthodox code choices like surface code, a modular universal fault-tolerant qLDPC architecture has been described and numerically benchmarked \citep{yoder2025tourgrossmodularquantum}. 


\section{\label{sec:level1} Cryogenic Systems}
\subsection{Optimized Temperatures and Cooling Powers for Cryostat Stages}

Many proposals for long-term fault-tolerant quantum computing require qubit numbers that are orders of magnitude greater than the capacity of available fridges, even when accounting for densification of components and reduction of thermal loads per qubit. One approach to extend fridges is to monolithically increase the size of the fridge. There are, however, challenges to increasing the size of a single volume. Practical concerns include limits of the facility capacities, for example, sizes of building entry points and load-bearing floor limits in typical data centers, while cost reduction per qubit is also critical and reduction of cost for equivalently scaled measures such as cooling power or working volumes have not yet been shown to improve through monolithic increase of fridge size. 

\begin{figure*}[] 
    \centering
    \includegraphics[height = 9 cm, width=\linewidth, clip,trim= 0mm 0mm 0mm 0mm]{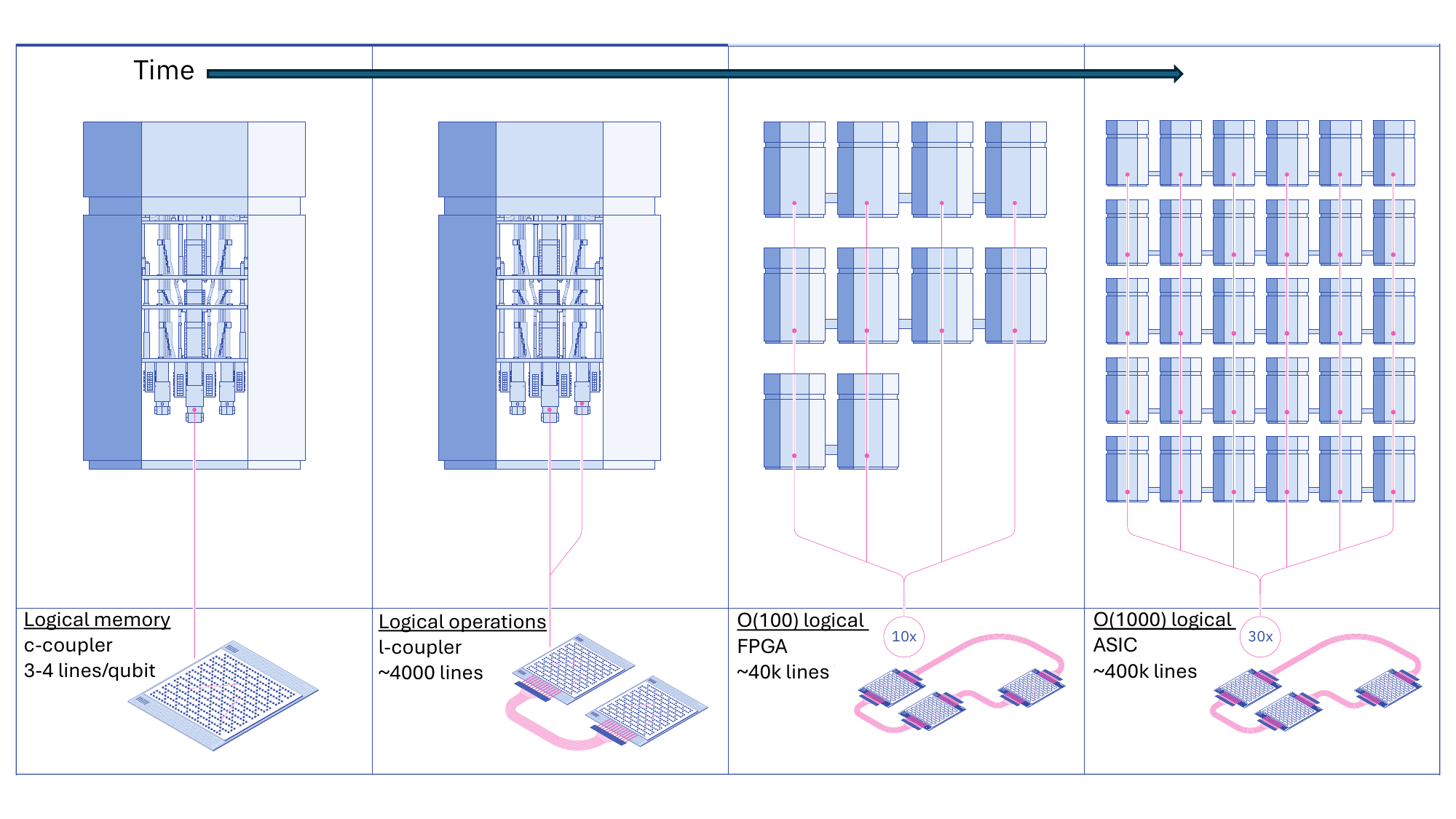}
    \caption{Example evolution of fridge extensibility to data center size. A contiguous volume is established through connection of fridge modules. Quantum processing chips are also modular connected using l-couplers as defined in Bravyi et al.\citep{bravyi_future_2022}. Commercial systems consisting of two dilution units, $\mathcal{O}$(25 $\mu$W) cooling power at the mixing chamber, and two to three pulse tubes has become a commonly used size and is a conceptual starting point for practical fridge module cooling powers, volume, weight, and cost. A third PT or some other source of cooling power would be needed for the elec.~cntrl.~stage.}
    \label{fig:extenFridge}
\end{figure*}

Modular fridge architectures, coupling unit fridges through tunnels \citep{magnard_microwave_2020}, for example, has been demonstrated in custom cases and represent an alternative path towards extensibility. Notionally, modular units that could be connected according to demand would allow growth of cryogenic environments according to need, while keeping cost per cooling power and volume at least relatively constant, while numbers of qubits per fridge might be scaled up (i.e., reducing cost per qubit), Fig.  \ref{fig:extenFridge}. Striving for minimum distances between adjacent modular fridge quantum processor payloads will also be desirable to maximize high-fidelity long-range coupling (l-coupling) \citep{bravyi_future_2022}. 

A sense of standardization, for example, of plate spacings and port sizes would also behoove the community, allowing fridge companies to focus on competitive advantage in cooling performance and cost while not being hampered by customization of retrofitting mechanical interfaces for componentry (e.g., cable and connector dimensions). Viewed from the perspective of the QC integrator, there are non-trivial time and cost barriers to retrofit a QC system to different mechanical configurations. To date, no fridge vendor offers a truly incrementally extensible modular solution.

\subsection{Power}
There are a number of dilution refrigerator vendors which offer similar `large' cooling power that have become workhorse units for QC integrators. These can be representative of a fridge module. Their capabilities center on a combination of 2-3 pulse tubes (PT) and dilution unit capability providing $\sim$20-30 $\mu$W of cooling power at 20 mK. Temperature zones generally breakdown into a PT1, PT2, still, cold plate (CP) and mixing chamber (MXC) stages, see Table \ref{table:coolPow}. We also assume that there will be an electronics control (elec. cntrl.) stage that supports cryoelectronics control (e.g., cryoCMOS) and that could operate at an intermediate temperature between 4-20 K. Assuming of the order 1,000 qubits per fridge unit at $\sim$1 mW per channel (i.e., 3-4 channels per qubit \citep{arute_quantum_2019-1,bravyi_high-threshold_2023}), the necessary cooling power for such a control electronics plate would be $\sim$30-40 W. Two pulse tubes applied to support a "large" fridge configuration is common and provides approximate power needs for the dilution system, radiative load, and passive wire cooling (see wiring section). A third PT or some other cooling power source would be necessary to supplying the tens of Watts for the electronics control, if the system design calls for moving a large fraction of control electronics into the fridge module. Coarse estimates of cooling powers for the different stages can be estimated from these order of magnitude considerations, see Table \ref{table:coolPow}.

\begin{table}[]
\begin{tabular}{||c | c | c ||} 
\hline
\makecell{Stage} &\makecell{Temperature (K)} &\makecell{Cooling power (W)} \\
\hline\hline
PT1     & $\leq$50 & 25 \\ 
 \hline

elec. cntrl     & $\sim$4-20 & 25-50 \\ 
 \hline

PT2     & $\sim$3.6-4.5 & $\sim$2-3 \\ 
 \hline

still     & $\leq$1.2 & $\sim$4-5$\times$10$^{-3}$  \\ 
 \hline
CP     & $\leq$0.2 & $\sim$1-2$\times$10$^{-4}$  \\ 
 \hline
MXC     & $\leq$0.02 & $\sim$2-3$\times$10$^{-5}$  \\ 
 \hline
\end{tabular}

\caption{Approximate magnitudes of available cooling powers needed for temperature stages of a modular fridge unit, see Fig. \ref{fig:extenFridge}, assuming roughly 2 PT and `large' fridge dilution unit cooling capacity assigned to all the stages other than the elec.~cntrl.~stage \citep{bluefors, maybell, oxford,zu_development_2022}. Some additional allowance on the higher end of the ranges is indicated for example for the PT2 stage to accommodate expected additional load of LNAs.}
\label{table:coolPow}
\end{table}

Wall power of scaled FTQC systems could grow to as large as GW scale and therefore become a relevant focus of design and innovation to minimize marginal run/compute cost. The fridge system wall plug power represents a large fraction of that power and alone could grow to comparable magnitudes as classic data centers when scaled to fault-tolerant error-corrected sizes. Pulse tubes are a dominant wall plug power demand, $\sim$10-15 kW with cooling power efficiencies of $\sim$1:1500, cooling power at the lowest temperature PT stage to wall plug power. The efficiency is sensitive to factors such as the operation temperature choice and optimization of thermal coupling between stage and PT (e.g., copper braid). Inclusion of cryogenic CMOS for control will further increase cooling power demands within the fridge with possibly $\sim$25 W to source $\sim$1000 of qubits and might suggest operation of stages at intermediate temperatures between $\sim$3.6 K and $\sim$50 K. As systems are scaled orders of magnitude, alternate higher efficiency cooling approaches will be desirable to provide `greener' and thereby cheaper computing. Efficiency can be expected to improve at higher temperature stages from simple arguments based on Carnot efficiency. Judicious placement of supporting componentry such as low-noise amplifiers (LNAs) and other electronics at as high a temperature as possible will be desirable. Cooling at higher temperatures furthermore provides more design space for alternative cooling technologies. There are also precedents of high $\sim$10 kW cooling using liquid helium cryoplants that for example provide efficiencies of $\sim$1:220 for $\sim$4.5 K \citep{shu_economics_2000}. However, there are nontrivial practical challenges for implementation including overcoming efficiency losses related to distribution and service-ability of future very large systems. In the context of National Laboratory experience in large scale cryogenic cooling, this may be an area of fruitful public-private collaboration.

\subsection{Rare resource utilization: $^{3}$He volume per qubit}
Dilution refrigerator technology is the predominant approach for cooling of continuous operation multi-qubit devices. Keeping the qubits as close to the achievable base operating temperature of these systems (e.g., targeting $\leq$20 mK), will become increasingly critical to minimize deleterious thermal excitations across increasingly large qubit numbers to achieve uniformly high-fidelity qubit operations and low leakage. Present mixtures of $^{3}$He/$^{4}$He are central to their operation. The $^{3}$He is extremely rare and has historically been dominantly supplied from the 12.3 year tritium half-life decay. The tritium supply is a byproduct of U.S. and Russian nuclear weapons stockpile stewardship. 

In the United States, the cost of $^3$He has been subsidized and is supplied through programs like isotopes.gov, which have mostly been sufficient for demand excepting historical demand spikes \citep{shea_helium-3_2020}. Notably, `at cost' production is estimated to be substantially higher particularly for alternate approaches like trace source extraction \citep{shea_helium-3_2020}. The U.S. and world production of $^3$He was for many years, $\mathcal{O}$($10^4$ liters/year). The majority of its use has been for purposes other than cryogenics related to quantum computing \citep{shea_helium-3_2020}. To address recent and anticipated greater future demand compared to supply, some companies (e.g., Air Liquide) have started to supply some of their $^3$He from niche stocks at commercial heavy-water electric generation reactors \citep{HeavyWaterReactor}. These reactors can capture and accumulate $^{3}$He. A source in Canada, for example, has been leveraged as a temporary buffer in response to demand and scarcity challenges.  

Scaling of mixing chamber power has relied primarily on incrementally adding dilution units to a fridge. Recently, IBM has shown that a 1000 qubit system $\sim$25 qubits/liter \citep{condor} is possible relying on dilution units that require $\sim$20 liters per unit. Systems of 100k qubits would demand large fractions of the yearly world production based on naive linear scaling. Planning for success, a combination of continued reduction of mixing chamber heat load per qubit, increased supply, and perhaps improvements in cooling power per liter $^3$He, will optimize the dilution refrigerator architecture with respect to cooling power vs temperature. There is a quadratic cooling power dependence on temperature for a dilution unit, and utilizing this effect when scaling the system has the promise of reducing the amount of $^3$He per qubit. Advances in alternative cooling approaches (e.g., continuous adiabatic demagnetization refrigeration) could complement $^3$He based cooling to avoid $^3$He becoming both a dominant cost per qubit and a potential supply bottleneck. 

 

\section{\label{sec:level1} Cryogenic Electronics}
\subsection{Testbeds for Cryogenic Electronics and Components, and Standardization Considerations} 

\label{sec:testbed}

In Ref.~\citep{boiko_low-noise_2023}, Boiko et al.~explain that: ``one of the limitations hindering a reliable QC supply chain is the ability for companies to provide microwave components that have been tested and qualified at cryogenic temperatures. However, the domains of cryogenics and microwave electronics are two highly specialized and rarely overlapping disciplines. The rapid increase in QC scaling drives the present need for dedicated test protocols and infrastructure for QC component testing...''

A new product family typically requires several design-fabricate-test cycles to reach market readiness. Currently, component manufacturers are equipped to develop and qualify their components in the standard temperature range of -40 °C to 85 °C. The cost of establishing cryogenic testing capability and the naturally longer test cycles prevent component manufacturers from fully qualifying their products for the QC market. Thus, the ability to perform economical cryogenic component testing is critical as QC systems scale and are commercialized. Why doesn’t this capability exist today? Two reasons stand out as most notable. First, there has not been significant business incentive from QC system integrators (i.e., those entities driving the scale and development of QC technologies) for commercially available test infrastructure, due in large part to the small-scale research nature of QC to date. Second, investment into the unique skill set and infrastructure is too costly for most companies.

Given the above, it would reduce barriers if Test as a Service (TaaS) could be incentivized to foster a healthy QC market ecosystem. Failing to establish this capability will result in several consequences that can impact the ability of the quantum industry to mature. For example, some of the standard temperature component manufacturers will find it prohibitively difficult to enter the QC market given their lack of ability to develop and test components that meet the requirements of system integrators. This will result in the component ecosystem being supported by only a small number of specialized component manufacturers. The absence of a competitive marketplace stymies both the innovations and the competitive labor market required to rapidly advance and scale cryogenic component technologies for QC. Volume testing will furthermore remain difficult and burdensome while standard methodologies are not available to unify the market or ensure accurate and repeatable product specifications. Traceable and certifiable testing at scale of components is also necessary to reach quality and reliability for production environments.

Example microwave components to be tested in a testbed include passive components such as RF wiring, connectors, qubit shielding, attenuators, filters, directional couplers, circulators and isolators, and active components such as LNAs, quantum-limited amplifiers (QLAs) and microwave switches. Characterization items differ slightly depending on the component and the level of integration. Required capabilities of a testbed are measurements of S-parameters, qubit performance, materials characteristics, failure in time (FIT) rates, thermal cycling reliability, etc. Scaling systems to $\mathcal{O}$(100k) qubits introduces a greater importance on reliability and FIT, particularly for components within the fridge, when considering the relatively high time overhead to warm and open the fridge environment to access and replace components.

There are a number of challenges to overcome in scaling the system to $\mathcal{O}$(100k) qubits, but one the first priorities will be to minimize the volume and weight of microwave passive components that incorporate permanent magnets, such as circulators and isolators, which are essential for routing the microwave pulse signals required to control and read out the qubits. For example, it has been reported that it is possible to reduce the volume to one-third and the weight to one-fifth compared to conventional isolators while maintaining high isolation characteristics over a wide frequency range. Approaches based on Josephson junctions are an appealing but speculative alternative worth further investigation as well. 
It seems, regardless, extremely challenging to pursue the plan illustrated in Fig.~1 without solving this problem (i.e., reducing volume and weight). If the microwave passive components are first miniaturized, the valuable large volume of the lowest temperature region of the dilution refrigerator can be effectively utilized to achieve quantum connections between modular processors as described in Bravyi et al.~\citep{bravyi_future_2022}.

Test components are placed inside a cryogenic refrigerator such as a dilution fridge, and the component is characterized, for example, the S-parameters are measured by a vector network analyzer. To increase throughput, multi-channel switches are a common feature allowing switching between multiple components within the fridge and smaller and faster switches are of interest. Further innovations providing faster cycle times and throughput of components are of great interest.

For the material characteristics, measurements such as thermal conductivity, resistance, etc.~and their temperature dependencies are performed. Characterization approaches depend critically on the type of component to be probed. Qubit performance tests include a wide variety of experiments to determine qubit device parameters, such as coherence times, gate errors, measurement errors, and error per layered gate (EPLG) experiments \citep{mckay_benchmarking_2023}. Whether one performs all or some of these experiments depends on whether the control and readout chain includes the component under test. For components on the readout line, for example, insertion loss and thermal isolation due to the component can be estimated from the impact on the readout SNR and qubit decoherence time (e.g., $T_2$), respectively. The component to be probed and a reference are arranged in parallel and switched with the multi-channel switch. Overall, qubit architecture specific measurements can be more challenging to provide as a service because of the level of customization and barriers to sharing of company information.

The need for a testbed to drive efficiency in the evaluation of, and to support standardization of cryogenic components is also a challenge facing the quantum design community in the context of deriving learning for custom CMOS designs (especially cryo-CMOS designs). It is a challenge to establish a representative testing context meaningful to test cryo-CMOS and it demands that researchers implement not only exploratory circuits, but also qubit test structures, connectivity solutions and a system framework. The development of one or more reference testbeds that (1) enable connection of exploratory circuit designs both to target qubit test payloads and to a system infrastructure that provides a wrapper to support qubit calibration and gate execution, and (2) support a means to compare to best-of-breed qubit interface electronics solutions would bring great value to the broader community.

In an attempt to meet this demand, for example, several testbeds for QC components are being planned in Japan by the National Institute of Advanced Industrial Science and Technology (G-QuAT) 
and Osaka University. 
Also, the University of Tokyo, in collaboration with IBM Quantum, has begun collaborating with Japanese companies to develop critical QC components that can function at dilution refrigerator temperatures, using testbeds installed at their Quantum Hardware Test Center.


\subsection{Cryogenic Wiring, Integration and Technology Foundries}

The i/o of signal between qubit and control electronics is a central bottleneck to scaling. Cost, thermal load and space must be scaled along with the number of qubits given a fixed fridge unit (i.e., fixed fridge volume, escape port sizes and cooling power) \citep{IEEE2023CryogenicElectronicsRoadmap,Krinner2019-ih,raicu2025cryogenicthermalmodelingmicrowave,manifold2025thermalcapacitymappingcryogenic}. Approaches to this challenge include multiplexing (e.g., readout and control), densification of the wiring (e.g., flex), introduction of control within the fridge (e.g., cryoCMOS) and alternative signal distribution approaches to modify bandwidth of channels and/or footprint (e.g., RF over fiber).

Flex is a promising path to substantially reducing cost, thermal load and space, while still obtaining sufficient performance (e.g., attenuation and crosstalk). Flex has been investigated for quantum computing \citep{tuckerman_flexible_2016,noauthor_osprey_nodate,noauthor_condor_nodate} but is, for QC, still at a relatively custom design and fabrication stage. There are furthermore few standards for quantum system integrators (e.g., lengths, thermal conductivity, attenuation, cross talk or connectorization choice) and the rapid need to scale drives rapidly changing specifications for a spectrum of quantum integrator custom solutions including the possibility that solutions will be eliminated in relatively short times as discrete elements are integrated into more compact and cost effective subsystems \citep{abdo_active_2019,das_4--6-ghz_2024}. This makes it extremely challenging for vendors to design their own solutions and intersect with quantum integrators because of challenges with cost of paying vendor margins for small market components, moving targets, timely communication and short lived solutions. 

A model of integrator design ownership (e.g., foundry model) is an alternative functional model that provides more agile response, while reducing risk for the manufacturer who can sell to multiple integrators avoiding custom in-house design costs and intersection challenges for what is a relatively small market. A need for vendors that can respond to both superconducting and normal metal flex fabrication is needed. Normal metal wiring will likely be a sustained need for at least the purpose of power supplies and some signaling even if other control electronics (e.g., cryoCMOS) is introduced at cold stages close to the quantum processor. There are fewer options to dislodge the need for superconducing wiring and therefore superconducting flex represents a continuing opportunity for both research and development of foundry-like options as improved materials and processing capability is needed. 

To further illustrate the parameter space, returning to the modular fridge model and superconducting wiring as an example. QC integrators will be driven towards 1000s of qubits per fridge-module due to drivers such as cost per qubit and practical use of space. Available space, exit port area, and cooling power of fridges can be roughly estimated at an order of magnitude as they are not very agile in view of practical challenges of cost to change, risk, and lead time to make significant changes to fridges. In this context, order of magnitude wiring density and thermal requirements can be made. If one assumes a high-degree qubit coupling architecture (e.g., $\sim$4 signals per qubit assuming tunable coupling \citep{bravyi_high-threshold_2023,arute_quantum_2019-1}) and a limited number of ISO100-like line-of-sight ports connecting to the superconducting lines, roughly $\sim$4-5 per fridge-module, densification of wiring might be assumed to push towards cabling line densities of $\geq$ 1 line/mm. This estimated density is driven in part due to an assumption of some minimum spacing between flex cables, limited due to practical considerations such as cable-to-cable connectorization limits leading to only space for $\mathcal{O}$(10) stacked cables of width $\mathcal{O}$(100) mm wide that pass the ISO100-like port or common side loading bay assembly sizes. Passive heat load management will further demand of the order 1 $\mu$W-cm/K assuming a length separation of $\sim$500 mm between a $\sim$4 K stage and the QPU. In contrast, estimates for signal performance requirements like bandwidths or tolerable crosstalk and loss are more challenging because of the qubit architecture specificity needed.  

In the context of QC integrators exploring many qubit architectures, the choice of cabling materials, stack-up and connectorization to simultaneously satisfy, mechanical, thermal, electrical and cost performance will likely continue to remain custom. In this light, having foundry capabilities for the flex technologies that can respond to custom requests appears to have a great deal of merit. 

Foundry capabilities apply well beyond the flex example. One might anticipate that CMOS foundries will play an ever greater role as ASIC solutions are developed (see later section), for example, including the displacement of present discrete components such as HEMT based LNAs by more integrated multi-channel and lower cost biCMOS or CMOS solutions \citep{bardin_cryogenic_2021}. In some qubit architectures, tolerance to lower measurement fidelity may allow performance to be traded for reductions in cost, size, weight and power (CSWaP). 

We also note a gap in superconducting foundry options that support development and commercialization of componentry that could significantly reduce CSWaP, such as, multi-channnel integrated isolators (e.g., non-ferrite approaches \citep{abdo_active_2019}), quantum-limited amplifiers and more speculatively low-power control electronics (e.g., DACs). CSWaP reduction or complete circumvention of readout components \citep{opremcak_high-fidelity_2021}, for example novel approaches to reduce isolator sizes \citep{mahoney_-chip_2017}, remain an active parallel and competing area of research and development.  

\begin{figure*}[] 
    \centering
    \includegraphics[width=0.9\linewidth, clip,trim= 0mm 0mm 0mm 0mm]{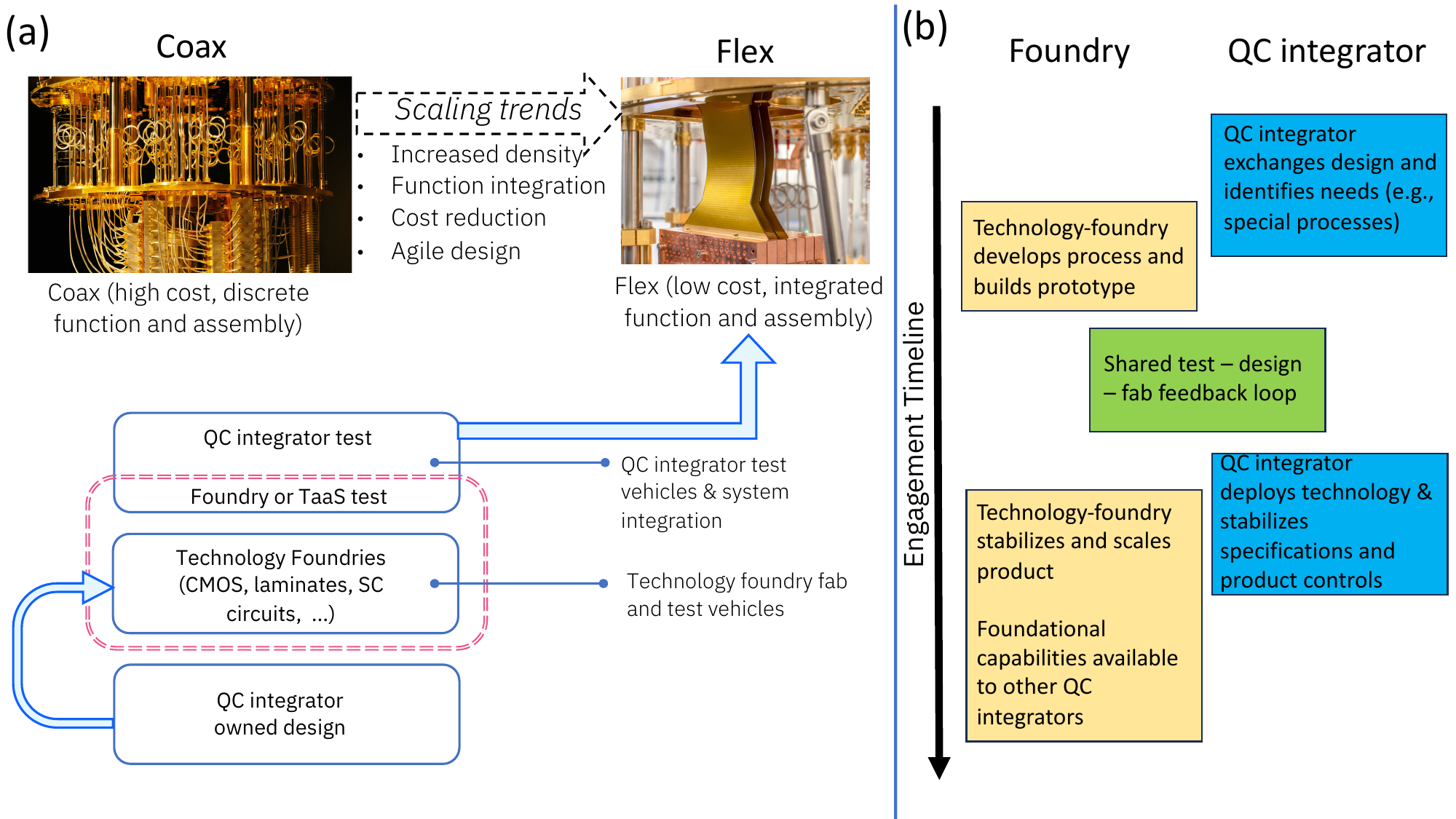}
    \caption{(a) Functional relationship and (b) conceptual engagement timeline of technology foundries, QC integrators and test services.}
    \label{fig:techfoundry}
\end{figure*}

We now turn to a notable emerging research and development thrust in superconducting cabling, which is driven by modular concepts entailing entanglement between separate chips within a shared cryogenic environment \citep{magnard_microwave_2020,niu_low-loss_2023}. The need for ultra-low-loss connectivity will introduce a need for specialized high-Q superconducting wiring with photon loss comparable to that of optical long haul fiber (e.g, order of 0.1-0.4 dB/km \citep{niu_low-loss_2023}) as well as the development of complementary low-loss connectorization to the QPU. The best results have been observed in Al cables with low density teflon dielectric. It will be of interest to identify other materials motivated by the desire to widen engineering options (e.g., bonding, cable handling) and this may introduce a need for improved basic understanding with respect to engineering RF loss in dielectrics and superconductors at the single excitation level in these cable/connector geometries including extending the communities understanding of defects and two level systems within the skin depth of the metals and within the cable dilectrics. The refinement of plating processess may be a fruitful direction. In the near term discrete cable connectivity will be sufficient, but longer term research and development of higher density scaled solutions to produce more seamless connection between remote chips will be also likely be needed. 

Another direction of interest is increasing bandwidth. More disruptive approaches such as RF over fiber \citep{lecocq_control_2021} are of academic interest in the context that they are high-risk high-reward paths. In the case of RF over fiber, it is unlikely that such solutions will reach the MXC and still be viable from a cooling power perspective. However, the substantial bandwidth and low thermal conductivity of fiber are tantalizing and might offer important options for spatial transposition of control electronics (i.e., further away from the fridge), while providing delivery of signal to higher temperature stages combined with perhaps fewer active components in the fridge than cryoCMOS. Space around the fridge for electronics is increasingly a scaling constraint as qubit number is scaled for a unit fridge footprint for which RF over fiber could be useful. However, whatever alternative solution is identified, reliability, thermal load and cost relative to relying on brute force wiring still need to be shown.  



\section{\label{sec:level1} Control Hardware}


\subsection{Functionality and Reliability}

Electronics solutions are needed to realize multiple functions in an error-corrected quantum computing system. These functions comprise RF control pulse generation, flux gate control signal generation, pump tone generation for quantum-limited amplifiers, RF readout pulse generation, low-noise amplification, readout discrimination, syndrome decoding, and system control and coordination. Electronics solutions must also be compatible with an end-to-end system design, software stack, and compiler infrastructure that enables appropriate abstraction to support efficient customer engagement with such systems. As systems increase in size and complexity, cryogenic electronic solutions may be needed to reduce signal travel times as well as wiring density. Current modalities envisioned for such control include cryogenic CMOS technology as well as electronic based on digital superconducting electronics. 

As systems scale, robustness and reliability grow in importance.  FIT rate analysis for the components used in quantum computing systems is therefore necessary.  For system elements that do or will operate at room temperature, FIT data exists or can be generated using well-understood methodologies.  For the elements of systems operating at cryogenic temperatures, however, FIT data does not exist and the appropriate conditions under which to acquire data and then develop associated models has not been defined. To address this gap, it is first necessary to identify intended use case classes for system components, including, for example, projected frequency of temperature cycling and target overall lifetimes.  Establishing standard approaches, for example,  to assess failure rates for cryogenic CMOS elements across multiple techonologies would also be valuable, as would the development of standardized approaches to evaluate failure rates for connectors, solder joins, flex cables, amplifiers, non-reciprocal elements, and other cryogenic system components. The creation of a test facility and an associated set of failure analysis methodologies that would drive clarity in the generation of failure rate performance would be very valuable in this context, see section \ref{sec:testbed}.

As described in more detail below, the choice of cryogenic CMOS versus room temperature CMOS for qubit interface electronics drives significant considerations regarding reliability, availability, serviceability, and performance.  As noted below, cryogenic CMOS usage promises to reduce connector count and wiring density into the dilution refrigerator, which could improve reliability and assembly, but introduces what is today a question mark, namely the relative reliability of CMOS and associated packaging at cryogenic temperatures. The reliability of cryogenic CMOS and associated packaging will also be a significant contributor to serviceability and availability for future systems.  Unlike systems that use room-temperature electronics for qubit interface control, systems that use cryogenic CMOS will require warm-up and cool-down if component replacement is required, making the need for a service strategy more critical, undoubtedly making servicing more challenging, and (comparatively) negatively impacting system availability. Key investments that would bring greater clarity to this space include reliability evaluation of cryogenic CMOS approaches and other cryogenic components as noted above.  As systems scale, even independent of specific reliability concerns associated with any one component, efforts to develop redundancy solutions that extend all the way to the qubit plane and/or error correction approaches that tolerate isolated hard failure of connections to a subset of system qubits would bring value. From a performance perspective, as described in greater detail below, the use of cryogenic elements imposes a  stricter power budget but offers potential for reduced noise and wiring complexity reduction; it also offers a path to support highly local feedback loops for branching operations, although the compelling value of such a capability versus loops closed through room temperature solutions is not presently obvious. 


\subsection{Technology Implementation}

Reduction of cost, size, weight and power is vital to achieving commerically viable systems. The path to further driving cost reduction necessarily demands the use of electronics solutions featuring levels of integration and customization (e.g., ASIC solutions) that are beyond those deployed in current known solutions.  For each electronic component, it is necessary to consider the physical placement in the system, both from a thermal environment perspective, and, for room temperature elements, with respect to the location of refrigerator electrical ports.

As noted in the cost discussion below, per-qubit electronics cost is a significant challenge that must be met as a necessary condition for the realization of quantum computing systems at scales envisioned to support quantum error correction. Current approaches \citep{ZettlesWJWA22} that rely on racks of custom boards built from commercially available electronic components are viable for systems with even up to 1000 qubits.  Such approaches, however, do not appear to support a path to the cost take-down needed for future systems as they scale toward qubit counts of order 100k and beyond. Integrated CMOS solutions operating at room temperature, within the cryostat, or in hybrid combinations are promising that would involve the exchange of a more significant up front non-recurring engineering cost and reduction in flexibility for a dramatic reduction in per-qubit bill-of-materials electronics cost. Extensive multiplexing has been proposed as a means to drive further cost reduction \citep{ParkIntel21}, but challenges are foreseen in effective support of quantum error correction in a highly multiplexed environment \citep{Frank23ISSCC}. As a room temperature alternative to a full room temperature ASIC implementation, the use of FPGAs (operating at room temperature) with integrated multi-channel DACs and ADCs as primary qubit interface electronics is worth exploration, but this approach appears unlikely to support electronics cost reduction to the degree needed for future scaled systems. Note, however, that in scaled quantum computing systems, FPGA usage as an element that provides a customizable interface to CMOS ASICs and supports critical computation tasks (e.g. syndrome matching) is a near certainty.

A room temperature ASIC approach offers multiple advantages over a cryogenic CMOS approach. Power consumption limits are significantly less stringent for a room temperature design than for a cryogenic design, opening up the design space and helping to mitigate traditional power/performance tradeoffs. Design for serviceability is also more straightforward as replacing a board housing a room temperature ASIC is unlikely to demand thermal cycling of the quantum computer's dilution refrigerator. Such an ASIC would also be designed within the supported temperature range for models and digital libraries associated with the vast majority of CMOS technology offerings of today. Such an approach also presents significant challenges compared to those associated with a cryogenic CMOS approach, however. Wiring density connecting elements at a cryogenic (e.g., 4K) stage of the dilution refrigerator to those at room temperature will be dramatically reduced if cryogenic CMOS is used. Similarly, connectorization challenges and connector density will be reduced in this case. Operating at lower temperature may further provide an opportunity for performance benefit in the cryogenic CMOS case, through reduction in thermal noise and improvement in other properties of CMOS technology at cryogenic temperatures. If superconducting flex can be used to link cryogenic CMOS elements to the qubit plane, an effective path to bypassing the signal quality/thermal conductivity trade space can be explored.

While cryogenic CMOS is promising approach,  technical investment on the part of the broader community in a number of areas would be highly beneficial to accelerating progress.  Achieving required performance while meeting cooling power constraints, especially if pulse tube coolers are used, is an area in which the research community has begun to make investments \citep{Yoo2024CryogenicCMOS,Chakraborty21,Kang2023CryogenicDRAG}, but significant further effort is required.  An additional high-level domain where investment is needed is in the area of CMOS technology and associated modeling.  Enabling advanced node reduced supply operation—-including for SRAM—-will drive better optimization and improved power efficiency  for cryogenic operation.  Device,  interconnect, and library modeling reflecting below 20 K behavior critical to enabling improved design, both to accurately predict design performance and to accurately predict design power consumption.  Improved thermal modeling and thermal solutions will let us better estimate and optimize controller temperature.  Understanding and mitigating stress effects for chip-package interactions at cryogenic temperatures will also be important. Reliability modeling for CMOS and other electronic components at cryogenic temperatures is needed to make scaled systems viable. Standardized approaches to enable serviceability of cryogenic CMOS will also be highly valuable. For both cryogenic and room temperature approaches, supporting investment in fully custom electronic designs despite the low volume nature of the current electronics for quantum computing market is a further imperative.


\subsection{Cost/Budget}

In the early days of the quantum program, quantum computing electronics were implemented primarily using off-the-shelf commercial instruments, resulting in an effective $\sim$\$50,000 per qubit for electronics cost alone. Some efforts at companies such as IBM have implemented custom discrete electronics-based solutions, which have targeted both dramatic densification of the electronics solution while also achieving a substantial improvement in cost per qubit (e.g., perhaps an order of magnitude). While addressing electronics cost is necessary to achieve an acceptable cost point for scaled quantum computing systems, it is not sufficient. Major fixed cost contributions to the cost per qubit roughly break down into the categories of electronics, wiring and cryogenic environment. Approaches to reduction of cost in these areas are noted in the sections above.

A general challenge that the community faces is that quantum computing is presently a low volume application. Any consideration of the implementation trade space must strongly reflect non-recurring engineering cost in addition to unit cost and cost of operation. Physical footprint at the system level may have significant capital infrastructure implications, especially as cryogenic environment sizes grow (i.e., multi-fridge-module systems). The physical footprint of elements to be placed at room temperature but ideally near refrigeration units is also significant in that floor space is intrinsically expensive, as is the cabling needed to connect electronics to fridge ports while meeting required signal integrity specifications. Finally, for cryogenic elements, the volume within the refrigerator is a constrained resource and is therefore drives cost considerations.   

It is also important to consider how use-case requirements will impact system cost. Current system designs are typically developed with flexibility and the need to support research in mind, the latter including not only qubit payload and gate exploration but also different means of implementing required control, filtering, and qubit interface functions. Scaled systems may eventually abandon the generality of today's research-focused designs in favor of more per-qubit cost-efficient and focused design points. This type of transition is important to enable the introduction of custom electronic designs into future quantum computing implementations, as ASICs typically will not deliver the same flexibility as will FPGAs or test equipment. Analysis and validation in the system-level and application context are critical to studying whether and under what circumstances lower performance, lower flexibility and lower cost elements might be appropriate to replace high performance, more expensive elements. One near term example might be, for example, determining alternatives to the high performance ultra low-noise amplifiers that are typically used in today's systems but for which readout performance might be sacrificed for certain qubit architectures and error correction choices.


\section{\label{sec:level1} QPU Tune-up and Operation}

At the highest level of the computational stack, one step up from the classical control hardware layer, we now identify challenges around QPU operation, including pulse-design, periodic re-calibration, error minimization strategies including error suppression, error mitigaton and error correction, algorithm implementation requirements, and ultimately, efficient validation  of the quantum output for quality assurance. All of these tasks involve significant complexity and some require significant resources. Below we summarize current best approaches and address critical needs where appropriate, which broadly points to mitigate the challenging scope of low-latency, high-performance classical compute that is required for many of these tasks. 

A major component of this section is various approaches to error minimization and performance validation, which are critical to optimize and verify the correctness of the QPU output.  As described below, there are a wide variety of error minimization strategies with distinct capabilities and requirements. At the lowest level there are passive run-time strategies such as dynamical decoupling and randomized compiling. At the next level of complexity there are error mitigation strategies requiring an expanded family of circuits as well as preparatory benchmarking experiments, verification \& validation of noise models and resource-intensive post-processing of data. Going forward fault-tolerant error correction will be adopted increasingly in the work loads. The computational load of these tasks of course depends on the details of the quantum algorithm, the error management strategy including architectural decisions of how to handle integration with high performance computing resources   \citep{kanazawa2025observabilityarchitecturequantumcentricsupercomputing}. 
  
\subsection{QPU Pulse-design, Tune-Up and Calibration}

    Preparing a quantum processor to execute an algorithm requires many parameters to be set precisely to allow high-fidelity gate execution (e.g., operation fidelities $>$ 0.999) with minimal cross-talk. Typical errors that must be minimized include information/coherence loss to the environment via, e.g., materials imperfections, such as drift and fluctuations, in the classical control fields used to operate the QPU. Achieving this aim requires procedures for pulse-design and calibration, as well as routinely scheduled (ideally at lower-cost) recalibration and tune-up procedures. As the system size grows, the number of measurements/calibrations associated with these tasks grows rapidly, highlighting the critical need for robust, automated and resource-efficient solutions for these tasks. These tasks include low-level error diagnostic routines, for example, to assess and reduce infidelities of gate and measurement operations, cross-talk and drift-compensation. Most of these techniques are well-established and standardized and require minimal classical compute resources. At the other end of the spectrum there are the more recently developed high-level, high-resource cost characterization methods, such as Pauli error reconstruction, which can characterize QPU clock cycles (parallel instruction sets) across very large arrays of qubits and identify cross-talk and correlations in multi-qubit error models. These methods have proven useful to characterize error models for large blocks of operations that are tremendously useful for error mitigation methods. They may also help optimize and inform decoding for quantum error correction, with the cost of requiring a significant fraction of QPU cycles and wall-clock time. A considerable amount of classical compute is needed even at current system sizes.

The more well-tested and well-established methods at the level of individual qubits and elementary operations include:
    \begin{itemize}
        \item Determination of the individual qubit frequencies and drive parameters for single and two-qubit gates via standard methods.
        \item Determination of the many-body cross-couplings in a given quantum processor graph via standard methods.
        \item Optimizing and benchmarking fidelities of a given pulse-design for sets of primitive gate operations is typically performed via randomized benchmarking (RB) \citep{emerson2005scalable} with random Clifford gates \citep{dankert2009exact,magesan2011scalable,magesan2012characterizing} or cross-entropy benchmarking \citep{XEB}, while  estimating the fidelity of individual entangling gates is generally achieved via cycle benchmarking (CB) \citep{erhard2019characterizing} with random Pauli for Clifford gates or dihedral gates \citep{Dugas_dihedral} for T-gates.
        \item More detailed (Pauli) error learning for quantum operations on small numbers of qubits is possible with slightly more resource-intensive approaches such as gate-set tomography \citep{GST2009} or cycle error reconstruction \citep{carignandugas2023error}, which combines the Pauli error learning ideas from \citep{Emerson2007} and the Pauli error amplification approach of CB \citep{erhard2019characterizing}. However, when the errors are quasi-local, they can be learned efficiently \citep{van_den_Berg_2023} and self-consistently \citep{chen2025disambiguatingpaulinoisequantum,chen2025efficientselfconsistentlearninggate} by adding constrained to the learning algorithm. 
        \item Optimizing and benchmarking fidelities of mid-circuit measurement and dynamic circuit operations using RB methods \citep{Govia2023MidCircuitRB,Shirizly2024DynamicRB, hothem2024measuring}, which are important for QEC and other circuit applications \citep{B_umer_2024,B_umer_2025,Kang_2025,Carrera_Vazquez_2024}   
        \item Older approaches to randomized benchmarking for elementary gates that are now less common include single qubit RB using a subset of the single Clifford gates \citep{Knill_2008}, which has been shown to produce a biased estimator \citep{Boone_2019}, randomized benchmarking with Haar random unitaries \citep{emerson2005scalable} which becomes inefficient for large numbers of qubits, and interleaved randomized benchmarking \citep{magesan2012efficient} with random Clifford twirls which shows significant bias due to systematic uncertainties in the presence of coherent errors \citep{Carignan_Dugas_2019a,Carignan_Dugas_2019b,Sannamoth2025}.
\item For more details, the reader is referred to the general theoretical  framework for randomized benchmarking developed in \citep{Helsen_2022}.
    \end{itemize}

   Methods for characterizing errors on large (parallel) sets of quantum operations, e.g., cycles, layers or large circuit blocks, have been less widely adopted, and can be resource intensive, both in terms of QPU time and classical post-processing. These methods include:
    \begin{itemize}
    
        \item Optimizing fidelity for layers or cycles of quantum gates across n-qubits, i.e., parallel instruction sets, in a manner that detects all cross-talk and context-dependent errors, can be achieved via cycle benchmarking (CB) \citep{erhard2019characterizing}, with a resource requirement that is constant in the number of qubits but also requires high-fidelity high-weight Pauli measurements which proves challenging in many platforms. Alternate methods such as simultaneous RB \citep{SimRB}, as simultaneous direct RB (i.e., layer fidelity (LF)) \citep{EPLG}, avoid measuring high-weight Pauli operators and have different resource requirements than CB with the drawback that although LF is a lower bound and often, to good approximation the fidelity, it is not exact in the presence of inter-gate cross-talk, see e.g.~ \citep{EPLG}. All of these approaches become more resource-efficient if low-latency, single-shot randomization is available in the control hardware \citep{HERC}.

        \item Diagnosing multi-qubit error rates, including cross-talk and context-dependent errors, is known to be possible in principle even across large number of qubits via Pauli error channel estimation techniques \citep{Emerson2007} and more scalable RB approaches \citep{hines2024fully}. Moreover the Pauli error rates can even be learned to multiplicative precision, for example, via the amplification approach of cycle error reconstruction (CER) \citep{erhard2019characterizing,carignandugas2023error,hashim_randomized_2020,chen2023learnability,fazio2025}, which has been extended to include mid-circuit measurements \citep{CERforMCM, CERforMCMb,}. In practice, the high QPU-time and classical post-processing required by Pauli error channel estimation imposes a  limitation on how much information can be learned (there is afterall an exponential amount of information available).  There is some evidence that estimation of even a  small fraction of Pauli error rates can predict logical error rates and even improve decoder performance \citep{fazio2025,Iyer2025}. Clifford benchmarking is also sufficient for some noise model classes \citep{merkel2025clifford}. Average circuit eigenvalue sampling \citep{hockings2024scalable} is an alternative scheme for Pauli error learning that is restricted to Clifford circuits, such as syndrome extraction circuits, but is known to give inaccurate results in the presence of inter-gate cross-talk. All of these schemes can have very high-resource costs and, as discussed below, more resource-efficient methods, based on validated error model assumptions, remain an important need and  area for further research. These methods also benefit significantly from low-latency single-shot randomization \citep{HERC}. This is especially true in the context of frequent, periodic recalibration.
        \item For more details, the reader is referred to a recent review article \citep{Hashim_2025}.
    \end{itemize}

While many of the above optimization tasks are resource intensive when performed comprehensively (i.e., without strong assumptions on the unknown error model), there is an important need to identify  short-cut methods to be used during the fequent, periodic recalibrations required during operation. As noted above, these short-cut methods may require additional, strong and system-specific assumptions on the error model, ideally validated by prior, more comprehensive error diagnostics, to reduce the scope of error learning \citep{harper_fast_2021}. Because all of the best-known methods for Pauli error learning require some form of twirling/randomization to induce a stochastic (Pauli) error channel,  the efficiency in terms of wall-clock time will certainly benefit greatly from low-latency single-shot randomization \citep{HERC}.  These short-cut methods could also benefit from using stronger twirling groups, such as single qubit or multi-qubit Cliffords instead of Pauli twirling \citep{Emerson2007}. Stronger twirls generally induce coarse-grained degeneracies in the learnable errors \citep{Emerson2007} and increase systematic uncertainty \citep{Carignan_Dugas_2019a,Carignan_Dugas_2019b}, so less information is learned and it is learned less accurately, but such methods can hope to learn these coarse-grained error properties much more efficiently. 



 Some other outstanding challenges include:
\begin{itemize}        
         \item All of the above ``established" methods make the standing assumption that non-Markovian errors are negligible. These effects can be important and so characterization of temporal fluctuations and drift is needed. While many of these effects show up as a deviation from an exponential decay in an RB experiment \citep{magesan2012characterizing,Wallman_2014,Ceasura2022}, the classification  of relevant non-Markovian effects and methods for their characterization remains an important open problem. 
         \item  There is an outstanding question as to whether the above methods will remain practical in the context of logical operations on logical (error-corrected) qubits, in particular whether the  standing Markovianity assumption will remain valid for such ``highly-engineered" qubits and gate operations.
        \item Optimizing / cross-layer compilation, and compiling to the actual implemented gates vs target gates, can provide significant advantages \citep{Foxen_2020, javadiabhari2024quantumcomputingqiskit} and requires further exploration.
        \item Optimizing selection of reduced instruction sets, by balancing trade-offs between calibration costs and compilation efficiency is also an opportunity, where the analysis of these trade-offs is application-specific and the possibility of a general framework remains an open question. An example implementation is described in \citep{javadiabhari2024quantumcomputingqiskit}). 
    \end{itemize}

\subsection{Error Suppression Methods During Operation}
  
  This is a broad category of strategies designed to suppress or correct errors arising in the course of QPU operation. These errors can include coherent errors due to  the finite precision and fluctuations/drift in the classical control fields and the decoherence errors due the environment.
  
  The family of solutions includes passive error suppression strategies such as dynamical decoupling (DD) \citep{Viola_1999}, Pauli frame randomization (PFR) \citep{Knill_2005} for Clifford circuits, and randomized compiling (RC) \citep{wallman2016noise} for universal circuits.
  
 Dynamical decoupling applies when errors are preferentially aligned along the quantization axis and  requires inserting additional gates to reduce/cancel these errors by leveraging the known (directional) bias in the error model \citep{Viola_1999}.  Pauli-frame randomization (PFR)  and  the equivalent but less well-known proposal known as PAREC \citep{Kern_2005}, which both require Pauli frame tracking (PFT) \citep{Knill_2005,WarePFR2021}, are designed to suppress unknown (but otherwise generic) coherent errors but are limited to the restricted setting of Clifford circuits due to the exponential cost of Pauli frame tracking through universal circuits. 
  
  Because of the limitation of DD to single-axis error, and restriction of PFR  to Clifford circuits, error suppression for universal circuits can be achieved via randomized compiling (RC) \citep{wallman2016noise}, a randomization method that leverages a locally corrected twirl and does not require Pauli frame tracking through the circuit. RC  utilizes either Pauli or dihedral twirls \citep{Dugas_dihedral}, depending on the how the T-gate is implemented. As a result, RC can be applied efficiently in the more general setting of universal circuits to suppress fidelity loss due to coherent errors \citep{wallman2016noise,hashim_randomized_2020,Ville_2022}. Randomized compiling also aligns the actual error model in applications, whether NISQ or logical level algorithms, with the error model that can be learned via Pauli error estimation methods discussed above \citep{wallman2016noise, carignandugas2023error}.  
  
  Randomized compiling is highly resource efficient. In terms of demands on QPU time, only 20 randomized circuits are required for significant suppression of coherent errors in many use-cases \citep{wallman2016noise,hashim_randomized_2020,Ville_2022}. It should be noted that this form of error suppression method does not require additional qubits nor access to low-latency high-performance compute, as is required by quantum error correction. It also does not require the exponential overhead of implementing a family of increasingly longer depth circuits, as is required for error mitigation. Both of these issues are discussed below.  
 However, for optimal resource-efficiency (in terms of wall-clock time) for suppression of (residual) coherent errors, RC works best by randomizing the single qubits gate in a circuit as frequently as possible. The randomization of the circuit should be achieved with zero or negligible latency, and the ideal limit is a fresh randomization with every new shot, which is already the current state of the art \citep{HERC}.

\subsection{Error Correction Methods During Operation}
   
Reaching the very low error rates required for many applications requires the implementation of fault-tolerance-quantum error correction (FT-QEC) strategies \citep{GottesmanPhysRevA,Gottesman2009introduction}, where the overhead of (many) additional qubits is leveraged to deliver lower error rates  at the level of logical qubits. This overhead is expected to take us to a performance regime where QPUs can deliver widespread quantum advantage. In typical implementations, FT-QEC requires encoding a small number of logical qubits into a much larger number of  physical qubits, and repeated measurement of error syndromes, for example as often as once per clock cycle. With each measured syndrome, the challenge is to quickly compute and implementing the most likely recovery operation --- this is a highly intensive task both in terms of repeated measurements, low-latency and high performance classical compute for the decoding of the measured syndromes --- and finally implementation of the computed recovery operation in the subsequent clock cycle. The qubit overheads for the best FT-QEC schemes are daunting, and the implementation of decoding on the time-scale of the quantum gates (clock cycle) remains an open challenge in many leading hardware platforms.
The performance and threshold of most FT-QEC is only well understood in the limit of stochastic error channels, such as the overly-simplistic  depolarizing channel or in some cases the more general setting of Pauli error channels which can of course be realized via randomized compiling \citep{wallman2016noise} 
as noted above and demonstrated experimentally in multiple platforms \citep{hashim_randomized_2020,fazio2025}.

Specific areas where significant advances are needed include:

 \begin{itemize}

\item  Control with very low latency i/o to the fpga/control hardware to deliver pre-computed randomized circuits, or else computing the randomizations on the fly at the fpga level \citep{HERC}, to tailor generic errors into stochastic Pauli errors  \citep{wallman2016noise} in the limit of single-shot randomization to minimize QPU wall-clock time. This is required also for Pauli error learing to predict and inform decoder performance.

    \item The implementation of FT-QEC requires developing new, architecture-friendly, low-overhead codes admitting fast and efficient decoders to avoid the extremely high qubit overheads demanded by, for example, the surface code. Recent progress on qLDPC codes \citep{bravyi_high-threshold_2023} suggests more research is needed to develop good qLDPC decoders in light of some promising recent results \citep{muller2025improvedbeliefpropagationsufficient, Roffe_PhysRevResearch.2.043423} and in parallel the community needs to explore the development of novel codes with desirable properties. Solutions are needed also to address the non-triviality of logical operations on logical qubits with codes such as qLDPC \citep{he2025extractorsqldpcarchitecturesefficient}, that address the time overhead of the logical operations \citep{yoder2025tourgrossmodularquantum} and fault-tolerant circuits that permit reduced connectivity \citep{Shaw_2025} or defects \citep{auger_fault-tolerance_2017}. Codes should be tested not just for memory, but also for logical operations.
    
    \item Low-latency and high-performance classical compute, eg at the FPGA or ASIC level, is required to implement decoding on platform with fast gates \citep{Battistel2023RealTime, maurer2025realtimedecodinggrosscode}. 
    
    \item Advancing error diagnostic methods to guide and inform code selection and decoder optimizations \citep{carignandugas2023error,fazio2025,hockings2024scalable,iyer2025enhancing}.
    
    \item Developing new error diagnostic methods relevant to the specific constraints of modular architectures.

    \item  Evaluating performance trade-offs and achievable physical and logical limits of error reduction through dynamical decoupling and randomized compiling

     \item Exploring decreased overheads in code design and decoder efficiencies that can be leveraged by qudit systems.

 \end{itemize}

\subsection{Error Mitigation via Expanded Circuits, Post-processing and Extrapolation} 

In recent years there has been significant interest in error mitigation methods that require executing an expanded family of (non-equivalent) circuits and post processing \citep{BabbushEM}. It is now increasingly accepted that these techniques could enable accurate observable estimation at qubit counts exceeding brute-force classical simulation, and with circuit volumes of few thousands of two-qubit gates, enabling initial demonstrations of quantum advantage \citep{eisert2025mindgapsfraughtroad,zimborás2025mythsquantumcomputationfault,aharonov2025importanceerrormitigationquantum}.  Furthermore, these techniques are also expected to be relevant even as error corrected processors become a reality. For example, extrapolation based methods, use circuits of increasing depth \citep{Dumitrescu2018CloudQC} or gate time \citep{kandala_error_2019} to artificially increase the noise, and then perform an extrapolation to the zero-noise limit to estimate the ideal output \citep{temme_error_2017}.  More controlled ways to perform extrapolation employ the use of a noise model. Such an approach requires measuring the error model for each cycle, eg via cycle error reconstruction, and post-processing leading to evidence of utility \citep{Kim2023QuantumUtility}. While error extrapolation techniques can retain a bias, given an accurate noise model, it is possible alternatively, to obtain an unbiased estimate for observables using Probabilistic Error Cancellation \citep{van_den_Berg_2023}. Generally, these approaches require exponential sampling overhead on the QPU, but in the limit of low error rates, the base of this exponential is very close to 1 and one can obtain unbiased observable estimates on circuits with gate counts (5000-10000) that could be challenging for classical simulation \citep{erhard2019characterizing,eisert2025mindgapsfraughtroad}.

 
Some key challenges include:
\begin{itemize}

    \item While cycle benchmarking provides an efficient method to bound the overhead cost for some error mitigation methods, limitations to the effectiveness of cycle benchmarking for this task include limitations under high measurement error rates, which needs to be overcome.  

    \item  Recent error mitigation methods, such as noiseless output extrapolation, suggest that lower overhead post-processing methods of error mitigation may be achievable \citep{Ferracin_2024}. 

   \item While error mitigation can lower the impact of the infidelity of the physical qubits, enabling more powerful computations in the near-term, the synergies between error mitigation and  quantum error correction require further exploration.

   \item Heuristic error mitigation methods can be useful but typically do not provide rigorous bounds on the mitigated results. The use of methods with tight, rigorous bounds is particularly crucial, for placing trust in error mitigated quantum computation, at scales where a classical solution does not exist \citep{lanes2025frameworkquantumadvantage}.

     \end{itemize}

\subsection{Algorithm Implementation}

  At the most basic level, classical data has to be uploaded to the QPU and outputs have to retrieved and analyzed. This already imposes a constraint on any algorithms where this basic i/o can require a significant, if not exponential cost.
  
Many quantum algorithms require leveraging complementary compute requirements, for example,  quantum-classical hybrid-algorithms. These algorithms often require low latency high-performance compute in parallel with the implementation of the quantum circuits. 
  
  With many recent advances both in machine learning and in different hardware architectures (eg. graphical and tensor processing units), some QPU-related tasks may be well aligned with specific classical tools, and may also form the basis of future quantum-classical co-design of such data interfaces. Research into the utility of quantum-classical hybrid algorithms is ongoing \citep{RobledoMoreno2024Chemistry,yu2025quantumcentricalgorithmsamplebasedkrylov}.

\begin{figure*}[] 
    \centering
    \includegraphics[width=\linewidth, clip,trim= 0mm 0mm 0mm 0mm]{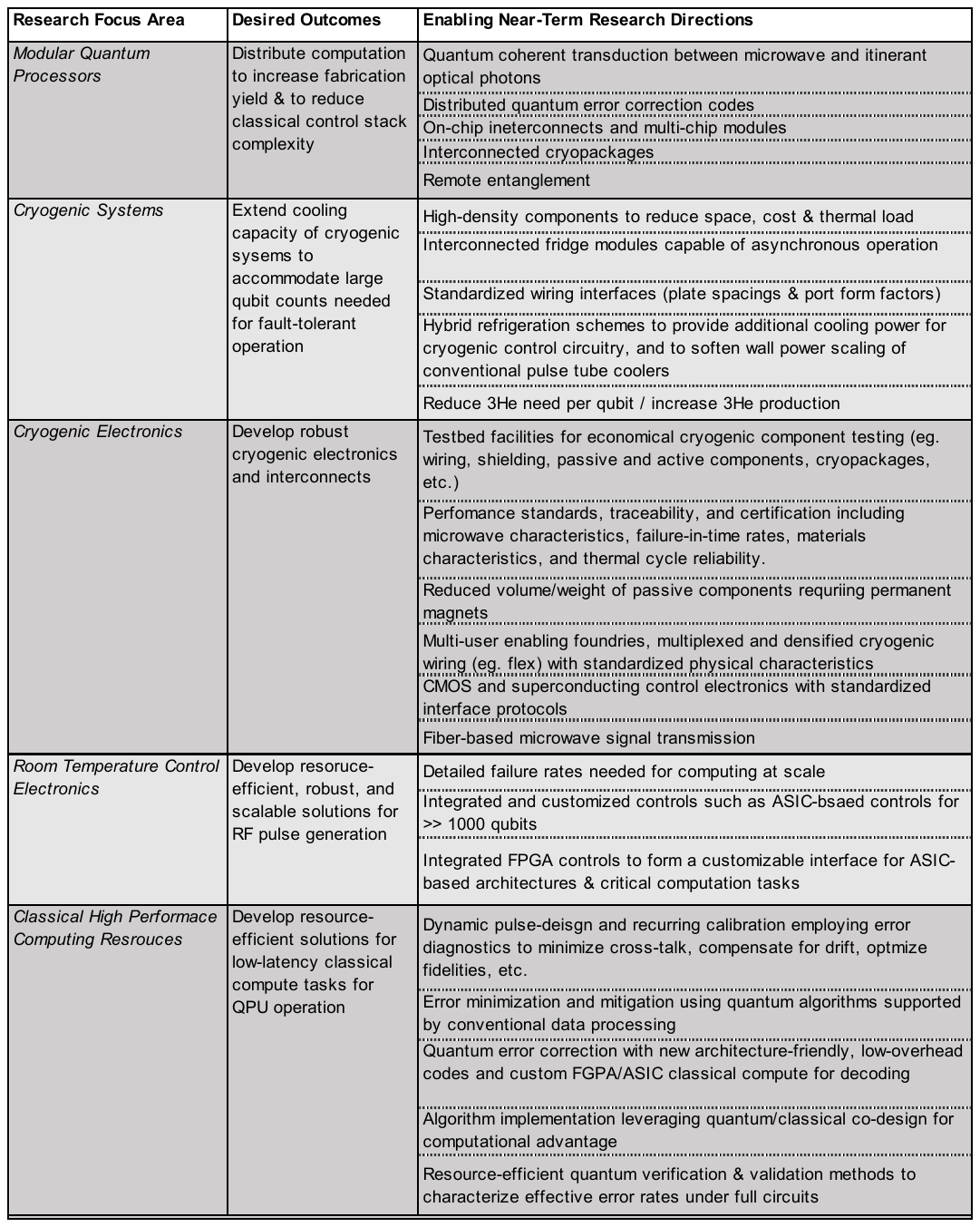}
    \caption{Summary of major recommendations}
    \label{fig:summary}
\end{figure*}


     

\subsection{Output Validation} 

      For many quantum applications and use-cases where an exponential advantage is expected, for example, quantum dynamical simulation, quantum chemistry, drug discovery and so on,  the correctness of the (candidate) quantum solution, obtained on imperfect and noisy quantum hardware, cannot be verified classically. In particular, the exponentially large number of unquantified residual physical errors create a problem where the precision or accuracy of the quantum solutions are unknown and can not be efficiently measured or efficiently computed.

    While it is well understood that this creates a validation problem for NISQ implementations of algorithms, because the presence of residual physical level errors lead to incorrect solutions, the persistence of this problem at the logical level is less well appreciated. It is important to recognize that fault-tolerant quantum computation is not actually ``tolerant  of all faults" and will always suffer from the same challenge. For example, the distance of the code guarantees only that a subset of (low-weight) errors are detectable and correctable, however, it is still possible to have larger weight errors go undetected or result in decoding errors (such as predicting the wrong recovery operation). 
     
    Thus, it is important to develop highly-scalable, resource-efficient verification and validation methods to characterize the effective error rates under all of the layers/cycles of universal circuit for large-scale processors, both for current NISQ processors and in the long-run of fault-tolerant architectures. This in turn provides a bound on the output accuracy under the entire algorithm when executed on given hardware. A promising solution in this direction consists in leveraging cycle benchmarking \citep{erhard2019characterizing} to measure the cycle infidelities for each of the distinct/layers of a quantum algorithm, then, thanks to the recent fidelity bound on compositions of cycles \citep{Carignan_Dugas_2019a}, this enables a rigorous and accurate `circuit benchmarking' \citep{QCAP} prediction of the circuit fidelity, even for universal circuits, and this circuit fidelity in turn bounds the total variation distance on the output. This approach assumes that the errors are stochastic, and this assumption can easily be enforced by implementing the circuit via randomized compiling.

    Going forward, some specific tasks and questions include:
    \begin{itemize}

     \item  Development of scalable verification and validation methods that can be applied at the logical level to herald or bound logical faults occurring due to code-distance limitations and decoding failures, such as the circuit benchmarking method. Determine how to related these bounds on fidelity and total variation distance to the observables of interest for quantum algorithms.  

     \item Can ideas from machine learning proved improved ways to learn noise/error models to ease scale-up?

\end{itemize}


\section{\label{sec:level1} Concluding Remarks}

Quantum information science and technology is a field that continues to grow at a rapid pace, and holds the promise of many new revolutionary technologies. To achieve the full potential of fault-tolerant quantum error corrected quantum computing, it will be vital to continue to improve the quality of individual qubits, produce quantum processors capable of generating long-lived entanglement, and develop robust, scalable strategies to address a QPU and interface it with a classical user. In this article, we have enumerated a list of several critical areas where advancements in the last category are needed for sustained future progress.  

We note that this progression is integrally connected with the establishment of a quantum ecosystem where many players in the commercial and academic domains can jointly develop quantum-tailored engineering solutions that require both fundamental science advances and industrial precision. At the heart of this environment is a quantum-literate workforce, and its cultivation cannot be understated. We envision this document as the first installment of a technology idea map, assembled by a collection of authors from across the ecosystem, for superconducting qubit technologies that is intended to help focus emerging technology partnerships. It will no doubt rapidly evolve with new developments in the field and should form the basis of a future living document.

\subsection{Acknowledgements}
The authors are grateful to the CIFAR Quantum Information Science program for their generous support of the roadmap workshop, and acknowledge insights and useful discussions with D. Lokken-Toyli, O. Dial, S. Engelmann, J. Orcutt, T. Yoder, S. Hart, L. Berge, A. Seif, A. Kandala, J. Raftery, M. Steffen and D. McKay. ANC acknowledges support from an ARO/LPS grant W911NF2310077 CGA acknowledges support from the EC through HORIZON-EIC-2022-PATHFINDEROPEN-01-101099697 (QUADRATURE). KS acknowledges support from the Center of Innovations for Sustainable Quantum AI (Japan Science and Technology Agency [JST], grant JPMJPF2221). AB acknowledges support from the Natural Science and Engineering Research Council of Canada.

\bibliography{lib_msc}

@Article{Michael2016,
  author    = {Michael, M. H. and Silveri, M. and Brierley, R. T. and Albert, V. V. and Salmilehto, J. and Jiang, Liang and Girvin , S. M. },
  journal   = {Physical Review X},
  title     = {New Class of Quantum Error-Correcting Codes for a Bosonic Mode},
  year      = {2016},
  number    = {031006},
  volume    = {6},
  doi       = {10.1103/PhysRevX.6.031006},
  publisher = {American Physical Society},
}

@Article{Gisin2007,
  author    = {Gisin, N. and Thew, R.},
  journal   = {Nature Photonics},
  title     = {Quantum communication},
  year      = {2007},
  pages     = {165--171},
  volume    = {1},
  doi       = {10.1038/nphoton.2007.22},
  publisher = {Springer Nature},
}

@article{kandala_error_2019,
	title = {Error mitigation extends the computational reach of a noisy quantum processor},
	volume = {567},
	issn = {1476-4687},
	url = {https://doi.org/10.1038/s41586-019-1040-7},
	doi = {10.1038/s41586-019-1040-7},
    year = {2019},
	pages = {491--495},
	number = {7749},
	journal = {Nature},
	shortjournal = {Nature},
	author = {Kandala, Abhinav and Temme, Kristan and Córcoles, Antonio D. and Mezzacapo, Antonio and Chow, Jerry M. and Gambetta, Jay M.}
}

@misc{hashim_randomized_2020,
	title = {Randomized compiling for scalable quantum computing on a noisy superconducting quantum processor},
	author = {Hashim, Akel and Naik, Ravi K. and Morvan, Alexis and Ville, Jean-Loup and Mitchell, Bradley and Kreikebaum, John Mark and Davis, Marc and Smith, Ethan and Iancu, Costin and O'Brien, Kevin P. and Hincks, Ian and Wallman, Joel J. and Emerson, Joseph and Siddiqi, Irfan},
	year = {2020},
	note = {\_eprint: 2010.00215},
}

@article{fowler_surface_2012,
	title = {Surface codes: Towards practical large-scale quantum computation},
	volume = {86},
	issn = {1050-2947, 1094-1622},
	url = {https://link.aps.org/doi/10.1103/PhysRevA.86.032324},
	doi = {10.1103/PhysRevA.86.032324},
	shorttitle = {Surface codes},
	pages = {032324},
	number = {3},
	journal = {Phys. Rev. A},
	author = {Fowler, Austin G. and Mariantoni, Matteo and Martinis, John M. and Cleland, Andrew N.},
	urldate = {2019-08-21},
	year = {2012-09},
	langid = {english},
}

@misc{gold_entanglement_2021,
	title = {Entanglement Across Separate Silicon Dies in a Modular Superconducting Qubit Device},
	author = {Gold, Alysson and Paquette, J. P. and Stockklauser, Anna and Reagor, Matthew J. and Alam, M. Sohaib and Bestwick, Andrew and Didier, Nicolas and Nersisyan, Ani and Oruc, Feyza and Razavi, Armin and Scharmann, Ben and Sete, Eyob A. and Sur, Biswajit and Venturelli, Davide and Winkleblack, Cody James and Wudarski, Filip and Harburn, Mike and Rigetti, Chad},
	year = {2021},
	note = {\_eprint: 2102.13293},
}

@article{zhong_deterministic_2021,
	title = {Deterministic multi-qubit entanglement in a quantum network},
	volume = {590},
	issn = {1476-4687},
	url = {http://dx.doi.org/10.1038/s41586-021-03288-7},
	doi = {10.1038/s41586-021-03288-7},
	pages = {571--575},
	number = {7847},
	journal = {Nature},
	author = {Zhong, Youpeng and Chang, Hung-Shen and Bienfait, Audrey and Dumur, Etienne and Chou, Ming-Han and Conner, Christopher R. and Grebel, Joel and Povey, Rhys G. and Yan, Haoxiong and Schuster, David I. and Cleland, Andrew N.},
	year = {2021-02},
	note = {Publisher: Springer Science and Business Media {LLC}},
}

@article{auger_fault-tolerance_2017,
	title = {Fault-tolerance thresholds for the surface code with fabrication errors},
	volume = {96},
	url = {https://doi.org/10.1103/PhysRevA.96.042316},
	doi = {10.1103/PhysRevA.96.042316},
	number = {4},
	journal = {Physical Review A},
	author = {Auger, James M.},
	year = {2017},
}

@article{temme_error_2017,
	title = {Error mitigation for short-depth quantum circuits},
	volume = {119},
	issn = {0031-9007, 1079-7114},
	url = {http://arxiv.org/abs/1612.02058},
	doi = {10.1103/PhysRevLett.119.180509},
	pages = {180509},
	number = {18},
	journal = {Physical Review Letters},
	shortjournal = {Phys. Rev. Lett.},
	author = {Temme, Kristan and Bravyi, Sergey and Gambetta, Jay M.},
	urldate = {2023-02-21},
	year = {2017-11-03},
	langid = {english},
	eprinttype = {arxiv},
	eprint = {1612.02058 [cond-mat, physics:quant-ph]},
}

@article{harper_fast_2021,
	title = {Fast Estimation of Sparse Quantum Noise},
	volume = {2},
	issn = {2691-3399},
	url = {https://link.aps.org/doi/10.1103/PRXQuantum.2.010322},
	doi = {10.1103/PRXQuantum.2.010322},
	pages = {010322},
	number = {1},
	journal = {{PRX} Quantum},
	shortjournal = {{PRX} Quantum},
	author = {Harper, Robin and Yu, Wenjun and Flammia, Steven T.},
	urldate = {2023-02-21},
	year = {2021-02-10},
	langid = {english},
}

@article{bravyi_future_2022,
	title = {The Future of Quantum Computing with Superconducting Qubits},
	volume = {132},
	issn = {0021-8979, 1089-7550},
	url = {http://arxiv.org/abs/2209.06841},
	doi = {10.1063/5.0082975},
	pages = {160902},
	number = {16},
	journal = {Journal of Applied Physics},
	shortjournal = {Journal of Applied Physics},
	author = {Bravyi, Sergey and Dial, Oliver and Gambetta, Jay M. and Gil, Dario and Nazario, Zaira},
	urldate = {2023-02-21},
	year = {2022-10-28},
	langid = {english},
	eprinttype = {arxiv},
	eprint = {2209.06841 [quant-ph]},
}

@article{boiko_low-noise_2023,
	title = {Low-noise amplifier cryogenic testbed validation in a {TaaS} (Testing-as-a-Service) framework},
	rights = {{arXiv}.org perpetual, non-exclusive license},
	url = {https://arxiv.org/abs/2309.03976},
	doi = {10.48550/ARXIV.2309.03976},
	author = {Boiko, Brandon and Zhang, Eric J. and Jorgesen, Doug and Engelmann, Sebastian and Grosskopf, Curtis and Paske, Ryan},
	urldate = {2023-10-04},
	year = {2023},
	note = {Publisher: {arXiv} Version Number: 1},
}

@article{tuckerman_flexible_2016,
	title = {Flexible superconducting Nb transmission lines on thin film polyimide for quantum computing applications},
	volume = {29},
	issn = {0953-2048, 1361-6668},
	url = {https://iopscience.iop.org/article/10.1088/0953-2048/29/8/084007},
	doi = {10.1088/0953-2048/29/8/084007},
	pages = {084007},
	number = {8},
	journal = {Superconductor Science and Technology},
	shortjournal = {Supercond. Sci. Technol.},
	author = {Tuckerman, David B and Hamilton, Michael C and Reilly, David J and Bai, Rujun and Hernandez, George A and Hornibrook, John M and Sellers, John A and Ellis, Charles D},
	urldate = {2023-10-11},
	year = {2016-08-01},
}

@article{magnard_microwave_2020,
	title = {Microwave Quantum Link between Superconducting Circuits Housed in Spatially Separated Cryogenic Systems},
	volume = {125},
	issn = {0031-9007, 1079-7114},
	url = {https://link.aps.org/doi/10.1103/PhysRevLett.125.260502},
	doi = {10.1103/PhysRevLett.125.260502},
	pages = {260502},
	number = {26},
	journal = {Physical Review Letters},
	shortjournal = {Phys. Rev. Lett.},
	author = {Magnard, P. and Storz, S. and Kurpiers, P. and Schär, J. and Marxer, F. and Lütolf, J. and Walter, T. and Besse, J.-C. and Gabureac, M. and Reuer, K. and Akin, A. and Royer, B. and Blais, A. and Wallraff, A.},
	urldate = {2023-10-11},
	year = {2020-12-21},
	langid = {english},
}

@article{niu_low-loss_2023,
	title = {Low-loss interconnects for modular superconducting quantum processors},
	volume = {6},
	issn = {2520-1131},
	url = {https://www.nature.com/articles/s41928-023-00925-z},
	doi = {10.1038/s41928-023-00925-z},
	pages = {235--241},
	number = {3},
	journal = {Nature Electronics},
	shortjournal = {Nat Electron},
	author = {Niu, Jingjing and Zhang, Libo and Liu, Yang and Qiu, Jiawei and Huang, Wenhui and Huang, Jiaxiang and Jia, Hao and Liu, Jiawei and Tao, Ziyu and Wei, Weiwei and Zhou, Yuxuan and Zou, Wanjing and Chen, Yuanzhen and Deng, Xiaowei and Deng, Xiuhao and Hu, Changkang and Hu, Ling and Li, Jian and Tan, Dian and Xu, Yuan and Yan, Fei and Yan, Tongxing and Liu, Song and Zhong, Youpeng and Cleland, Andrew N. and Yu, Dapeng},
	urldate = {2023-10-11},
	year = {2023-02-16},
	langid = {english},
}

@book{shea_helium-3_2020,
	title = {The helium-3 shortage : supply, demand, and options for Congress},
	url = {https://search.library.wisc.edu/catalog/9913067331802121},
	publisher = {[Library of Congress public edition]. [Washington, D.C.] : Congressional Research Service, 2020-},
	author = {Shea, Dana A. author},
	year = {2020},
}

@incollection{shu_economics_2000,
	location = {Boston, {MA}},
	title = {Economics of Large Helium Cryogenic Systems: Experience from Recent Projects at {CERN}},
	isbn = {978-1-4613-6892-2 978-1-4615-4215-5},
	url = {http://link.springer.com/10.1007/978-1-4615-4215-5_44},
	shorttitle = {Economics of Large Helium Cryogenic Systems},
	pages = {1301--1308},
	booktitle = {Advances in Cryogenic Engineering},
	publisher = {Springer {US}},
	author = {Claudet, S. and Gayet, Ph. and Lebrun, Ph. and Tavian, L. and Wagner, U.},
	editor = {Shu, Quan-Sheng},
	urldate = {2023-11-06},
	year = {2000},
	doi = {10.1007/978-1-4615-4215-5_44},
}

@article{mckay_benchmarking_2023,
	title = {Benchmarking Quantum Processor Performance at Scale},
	rights = {{arXiv}.org perpetual, non-exclusive license},
	url = {https://arxiv.org/abs/2311.05933},
	doi = {10.48550/ARXIV.2311.05933},
    year = {2023},
	author = {{McKay}, David C. and Hincks, Ian and Pritchett, Emily J. and Carroll, Malcolm and Govia, Luke C. G. and Merkel, Seth T.},
	urldate = {2024-03-08},
	year = {2023},
	note = {Publisher: [object Object]
Version Number: 1},
}

@article{das_4--6-ghz_2024,
	title = {A 4-to-6-{GHz} Cryogenic {CMOS} {LNA} With 4.4-K Average Noise Temperature in 22-nm {FDSOI}},
	issn = {2771-9588, 2771-957X},
	url = {https://ieeexplore.ieee.org/document/10423722/},
	doi = {10.1109/LMWT.2024.3355046},
	pages = {1--4},
	journal = {{IEEE} Microwave and Wireless Technology Letters},
	shortjournal = {{IEEE} Microw. Wireless Tech. Lett.},
	author = {Das, Sayan and Raman, Sanjay and Bardin, Joseph C.},
	urldate = {2024-03-09},
	year = {2024},
}

@article{abdo_active_2019,
	title = {Active protection of a superconducting qubit with an interferometric Josephson isolator},
	volume = {10},
	issn = {2041-1723},
	url = {https://www.nature.com/articles/s41467-019-11101-3},
	doi = {10.1038/s41467-019-11101-3},
    year = {2019},
	pages = {3154},
	number = {1},
	journal = {Nature Communications},
	shortjournal = {Nat Commun},
	author = {Abdo, Baleegh and Bronn, Nicholas T. and Jinka, Oblesh and Olivadese, Salvatore and Córcoles, Antonio D. and Adiga, Vivekananda P. and Brink, Markus and Lake, Russell E. and Wu, Xian and Pappas, David P. and Chow, Jerry M.},
	urldate = {2024-03-09},
	year = {2019-07-17},
	langid = {english},
}

@article{lecocq_control_2021,
	title = {Control and readout of a superconducting qubit using a photonic link},
	volume = {591},
	issn = {0028-0836, 1476-4687},
	url = {https://www.nature.com/articles/s41586-021-03268-x},
	doi = {10.1038/s41586-021-03268-x},
	pages = {575--579},
	number = {7851},
	journal = {Nature},
	shortjournal = {Nature},
	author = {Lecocq, F. and Quinlan, F. and Cicak, K. and Aumentado, J. and Diddams, S. A. and Teufel, J. D.},
	urldate = {2024-03-09},
	year = {2021-03-25},
	langid = {english},
}

@article{zu_development_2022,
	title = {Development of dilution refrigerators—A review},
	volume = {121},
	issn = {00112275},
	url = {https://linkinghub.elsevier.com/retrieve/pii/S001122752100148X},
	doi = {10.1016/j.cryogenics.2021.103390},
	pages = {103390},
	journal = {Cryogenics},
	shortjournal = {Cryogenics},
	author = {Zu, H. and Dai, W. and De Waele, A.T.A.M.},
	urldate = {2024-03-27},
	year = {2022-01},
	langid = {english},
}

@article{bravyi_high-threshold_2023,
	title = {High-threshold and low-overhead fault-tolerant quantum memory},
	rights = {{arXiv}.org perpetual, non-exclusive license},
	url = {https://arxiv.org/abs/2308.07915},
	doi = {10.48550/ARXIV.2308.07915},
	author = {Bravyi, Sergey and Cross, Andrew W. and Gambetta, Jay M. and Maslov, Dmitri and Rall, Patrick and Yoder, Theodore J.},
	urldate = {2024-03-27},
	year = {2023},
	note = {Publisher: [object Object]
Version Number: 2},
}

@article{bardin_cryogenic_2021,
	title = {Cryogenic Low-Noise Amplifiers: Noise Performance and Power Dissipation},
	volume = {13},
	rights = {https://ieeexplore.ieee.org/Xplorehelp/downloads/license-information/{IEEE}.html},
	issn = {1943-0582, 1943-0590},
	url = {https://ieeexplore.ieee.org/document/9467124/},
	doi = {10.1109/MSSC.2021.3072803},
	shorttitle = {Cryogenic Low-Noise Amplifiers},
	pages = {22--35},
	number = {2},
    year = {2021},
    journal = {{IEEE} Solid-State Circuits Magazine},
	journal = {{IEEE} Solid-State Circuits Magazine},
	shortjournal = {{IEEE} Solid-State Circuits Mag.},
	author = {Bardin, Joseph C.},
	urldate = {2024-03-27}
}

@article{opremcak_high-fidelity_2021,
	title = {High-Fidelity Measurement of a Superconducting Qubit Using an On-Chip Microwave Photon Counter},
	volume = {11},
	issn = {2160-3308},
	url = {https://link.aps.org/doi/10.1103/PhysRevX.11.011027},
	doi = {10.1103/PhysRevX.11.011027},
	pages = {011027},
	number = {1},
	journal = {Physical Review X},
	shortjournal = {Phys. Rev. X},
	author = {Opremcak, A. and Liu, C. H. and Wilen, C. and Okubo, K. and Christensen, B. G. and Sank, D. and White, T. C. and Vainsencher, A. and Giustina, M. and Megrant, A. and Burkett, B. and Plourde, B. L. T. and {McDermott}, R.},
	urldate = {2024-03-27},
	year = {2021-02-10},
	langid = {english},
}

@article{mahoney_-chip_2017,
	title = {On-Chip Microwave Quantum Hall Circulator},
	volume = {7},
	rights = {https://creativecommons.org/licenses/by/4.0/},
	issn = {2160-3308},
	url = {https://link.aps.org/doi/10.1103/PhysRevX.7.011007},
	doi = {10.1103/PhysRevX.7.011007},
	pages = {011007},
	number = {1},
	journal = {Physical Review X},
	shortjournal = {Phys. Rev. X},
	author = {Mahoney, A. C. and Colless, J. I. and Pauka, S. J. and Hornibrook, J. M. and Watson, J. D. and Gardner, G. C. and Manfra, M. J. and Doherty, A. C. and Reilly, D. J.},
	urldate = {2024-03-27},
	year = {2017-01-24},
	langid = {english},
}

@online{noauthor_condor_nodate,
	title = {Condor},
	url = {https://en.wikipedia.org/wiki/IBM_Condor},
}

@online{noauthor_osprey_nodate,
	title = {Osprey},
	url = {https://en.wikipedia.org/wiki/IBM_Osprey},
}

@article{arute_quantum_2019-1,
	title = {Quantum supremacy using a programmable superconducting processor},
	volume = {574},
	issn = {0028-0836, 1476-4687},
	url = {https://www.nature.com/articles/s41586-019-1666-5},
	doi = {10.1038/s41586-019-1666-5},
	pages = {505--510},
	number = {7779},
	journal = {Nature},
	shortjournal = {Nature},
	author = {Arute, Frank and Arya, Kunal and Babbush, Ryan and Bacon, Dave and Bardin, Joseph C. and Barends, Rami and Biswas, Rupak and Boixo, Sergio and Brandao, Fernando G. S. L. and Buell, David A. and Burkett, Brian and Chen, Yu and Chen, Zijun and Chiaro, Ben and Collins, Roberto and Courtney, William and Dunsworth, Andrew and Farhi, Edward and Foxen, Brooks and Fowler, Austin and Gidney, Craig and Giustina, Marissa and Graff, Rob and Guerin, Keith and Habegger, Steve and Harrigan, Matthew P. and Hartmann, Michael J. and Ho, Alan and Hoffmann, Markus and Huang, Trent and Humble, Travis S. and Isakov, Sergei V. and Jeffrey, Evan and Jiang, Zhang and Kafri, Dvir and Kechedzhi, Kostyantyn and Kelly, Julian and Klimov, Paul V. and Knysh, Sergey and Korotkov, Alexander and Kostritsa, Fedor and Landhuis, David and Lindmark, Mike and Lucero, Erik and Lyakh, Dmitry and Mandrà, Salvatore and {McClean}, Jarrod R. and {McEwen}, Matthew and Megrant, Anthony and Mi, Xiao and Michielsen, Kristel and Mohseni, Masoud and Mutus, Josh and Naaman, Ofer and Neeley, Matthew and Neill, Charles and Niu, Murphy Yuezhen and Ostby, Eric and Petukhov, Andre and Platt, John C. and Quintana, Chris and Rieffel, Eleanor G. and Roushan, Pedram and Rubin, Nicholas C. and Sank, Daniel and Satzinger, Kevin J. and Smelyanskiy, Vadim and Sung, Kevin J. and Trevithick, Matthew D. and Vainsencher, Amit and Villalonga, Benjamin and White, Theodore and Yao, Z. Jamie and Yeh, Ping and Zalcman, Adam and Neven, Hartmut and Martinis, John M.},
	urldate = {2024-03-28},
	year = {2019-10-24},
	langid = {english},
}

@article{QIU2024,
title = {Deterministic quantum state and gate teleportation between distant superconducting chips},
journal = {Science Bulletin},
year = {2024},
issn = {2095-9273},
doi = {https://doi.org/10.1016/j.scib.2024.11.047},
url = {https://www.sciencedirect.com/science/article/pii/S209592732400879X},
author = {Jiawei Qiu and Yang Liu and Ling Hu and Yukai Wu and Jingjing Niu and Libo Zhang and Wenhui Huang and Yuanzhen Chen and Jian Li and Song Liu and Youpeng Zhong and Luming Duan and Dapeng Yu}
}

@article{leung_deterministic_2019,
	title = {Deterministic bidirectional communication and remote entanglement generation between superconducting qubits},
	volume = {5},
	issn = {2056-6387},
	url = {https://www.nature.com/articles/s41534-019-0128-0},
	doi = {10.1038/s41534-019-0128-0},
	pages = {18},
	number = {1},
	journal = {npj Quantum Information},
	shortjournal = {npj Quantum Inf},
	author = {Leung, N. and Lu, Y. and Chakram, S. and Naik, R. K. and Earnest, N. and Ma, R. and Jacobs, K. and Cleland, A. N. and Schuster, D. I.},
	urldate = {2024-12-18},
	year = {2019-02-15},
	langid = {english},
}

@article{meesala_quantum_2024,
	title = {Quantum Entanglement between Optical and Microwave Photonic Qubits},
	volume = {14},
	issn = {2160-3308},
	url = {https://link.aps.org/doi/10.1103/PhysRevX.14.031055},
	doi = {10.1103/PhysRevX.14.031055},
	pages = {031055},
	number = {3},
	journal = {Physical Review X},
	shortjournal = {Phys. Rev. X},
	author = {Meesala, Srujan and Lake, David and Wood, Steven and Chiappina, Piero and Zhong, Changchun and Beyer, Andrew D. and Shaw, Matthew D. and Jiang, Liang and Painter, Oskar},
	urldate = {2024-12-18},
	year = {2024-09-30},
	langid = {english},
}

@article{kumar_quantum-enabled_2023,
	title = {Quantum-enabled millimetre wave to optical transduction using neutral atoms},
	volume = {615},
	issn = {0028-0836, 1476-4687},
	url = {https://www.nature.com/articles/s41586-023-05740-2},
	doi = {10.1038/s41586-023-05740-2},
	pages = {614--619},
	number = {7953},
	journal = {Nature},
	shortjournal = {Nature},
	author = {Kumar, Aishwarya and Suleymanzade, Aziza and Stone, Mark and Taneja, Lavanya and Anferov, Alexander and Schuster, David I. and Simon, Jonathan},
	urldate = {2024-12-19},
	year = {2023-03-23},
	langid = {english},
}

@article{fan_superconducting_2018,
	title = {Superconducting cavity electro-optics: A platform for coherent photon conversion between superconducting and photonic circuits},
	volume = {4},
	issn = {2375-2548},
	url = {https://www.science.org/doi/10.1126/sciadv.aar4994},
	doi = {10.1126/sciadv.aar4994},
	shorttitle = {Superconducting cavity electro-optics},
	pages = {eaar4994},
	number = {8},
	journal = {Science Advances},
	shortjournal = {Sci. Adv.},
	author = {Fan, Linran and Zou, Chang-Ling and Cheng, Risheng and Guo, Xiang and Han, Xu and Gong, Zheng and Wang, Sihao and Tang, Hong X.},
	urldate = {2024-12-19},
	year = {2018-08-03},
	langid = {english},
}

@article{xu_bidirectional_2021,
	title = {Bidirectional interconversion of microwave and light with thin-film lithium niobate},
	volume = {12},
	issn = {2041-1723},
	url = {https://www.nature.com/articles/s41467-021-24809-y},
	doi = {10.1038/s41467-021-24809-y},
	pages = {4453},
	number = {1},
	journal = {Nature Communications},
	shortjournal = {Nat Commun},
	author = {Xu, Yuntao and Sayem, Ayed Al and Fan, Linran and Zou, Chang-Ling and Wang, Sihao and Cheng, Risheng and Fu, Wei and Yang, Likai and Xu, Mingrui and Tang, Hong X.},
	urldate = {2024-12-19},
	year = {2021-07-22},
	langid = {english},
}

@article{sahu_quantum-enabled_2022,
	title = {Quantum-enabled operation of a microwave-optical interface},
	volume = {13},
	issn = {2041-1723},
	url = {https://www.nature.com/articles/s41467-022-28924-2},
	doi = {10.1038/s41467-022-28924-2},
	pages = {1276},
	number = {1},
	journal = {Nature Communications},
	shortjournal = {Nat Commun},
	author = {Sahu, Rishabh and Hease, William and Rueda, Alfredo and Arnold, Georg and Qiu, Liu and Fink, Johannes M.},
	urldate = {2024-12-19},
	year = {2022-03-11},
	langid = {english},
}

@article{higginbotham_harnessing_2018,
	title = {Harnessing electro-optic correlations in an efficient mechanical converter},
	volume = {14},
	issn = {1745-2473, 1745-2481},
	url = {https://www.nature.com/articles/s41567-018-0210-0},
	doi = {10.1038/s41567-018-0210-0},
	pages = {1038--1042},
	number = {10},
	journal = {Nature Physics},
	shortjournal = {Nature Phys},
	author = {Higginbotham, A. P. and Burns, P. S. and Urmey, M. D. and Peterson, R. W. and Kampel, N. S. and Brubaker, B. M. and Smith, G. and Lehnert, K. W. and Regal, C. A.},
	urldate = {2024-12-19},
	year = {2018-10},
	langid = {english},
}

@article{jiang_optically_2023,
	title = {Optically heralded microwave photon addition},
	volume = {19},
	issn = {1745-2473, 1745-2481},
	url = {https://www.nature.com/articles/s41567-023-02129-w},
	doi = {10.1038/s41567-023-02129-w},
	pages = {1423--1428},
	number = {10},
	journal = {Nature Physics},
	shortjournal = {Nat. Phys.},
	author = {Jiang, Wentao and Mayor, Felix M. and Malik, Sultan and Van Laer, Raphaël and {McKenna}, Timothy P. and Patel, Rishi N. and Witmer, Jeremy D. and Safavi-Naeini, Amir H.},
	urldate = {2024-12-19},
	year = {2023-10},
	langid = {english},
}

@article{zhao_electro-optic_2023,
	title = {Electro-optic transduction in silicon via gigahertz-frequency nanomechanics},
	volume = {10},
	issn = {2334-2536},
	url = {https://opg.optica.org/abstract.cfm?URI=optica-10-6-790},
	doi = {10.1364/OPTICA.479162},
	pages = {790},
	number = {6},
	journal = {Optica},
	shortjournal = {Optica},
	author = {Zhao, Han and Bozkurt, Alkim and Mirhosseini, Mohammad},
	urldate = {2024-12-19},
	year = {2023-06-20},
	langid = {english},
}

@article{ang_arquin_2024,
	title = {{ARQUIN}: Architectures for Multinode Superconducting Quantum Computers},
	volume = {5},
	issn = {2643-6809, 2643-6817},
	url = {https://dl.acm.org/doi/10.1145/3674151},
	doi = {10.1145/3674151},
	shorttitle = {{ARQUIN}},
    year = {2024},
	pages = {1--59},
	number = {3},
    journal = {{ACM} Transactions on Quantum Computing},
	journal = {{ACM} Transactions on Quantum Computing},
	shortjournal = {{ACM} Transactions on Quantum Computing},
	author = {Ang, James and Carini, Gabriella and Chen, Yanzhu and Chuang, Isaac and Demarco, Michael and Economou, Sophia and Eickbusch, Alec and Faraon, Andrei and Fu, Kai-Mei and Girvin, Steven and Hatridge, Michael and Houck, Andrew and Hilaire, Paul and Krsulich, Kevin and Li, Ang and Liu, Chenxu and Liu, Yuan and Martonosi, Margaret and {McKay}, David and Misewich, Jim and Ritter, Mark and Schoelkopf, Robert and Stein, Samuel and Sussman, Sara and Tang, Hong and Tang, Wei and Tomesh, Teague and Tubman, Norm and Wang, Chen and Wiebe, Nathan and Yao, Yongxin and Yost, Dillon and Zhou, Yiyu},
	urldate = {2024-12-19},
	year = {2024-09-30},
	langid = {english},
}

@article{storz_loophole-free_2023,
	title = {Loophole-free Bell inequality violation with superconducting circuits},
	volume = {617},
	issn = {0028-0836, 1476-4687},
	url = {https://www.nature.com/articles/s41586-023-05885-0},
	doi = {10.1038/s41586-023-05885-0},
	pages = {265--270},
	number = {7960},
	journal = {Nature},
	shortjournal = {Nature},
	author = {Storz, Simon and Schär, Josua and Kulikov, Anatoly and Magnard, Paul and Kurpiers, Philipp and Lütolf, Janis and Walter, Theo and Copetudo, Adrian and Reuer, Kevin and Akin, Abdulkadir and Besse, Jean-Claude and Gabureac, Mihai and Norris, Graham J. and Rosario, Andrés and Martin, Ferran and Martinez, José and Amaya, Waldimar and Mitchell, Morgan W. and Abellan, Carlos and Bancal, Jean-Daniel and Sangouard, Nicolas and Royer, Baptiste and Blais, Alexandre and Wallraff, Andreas},
	urldate = {2024-12-19},
	year = {2023-05-11},
	langid = {english},
}

@misc{bluefors, 
author = {{BlueFors}}, 
title = {BlueFors Website}, 
year = {2025}, 
url = {https://bluefors.com/} 
}

@misc{oxford, 
author = {{Oxford}}, 
title = {Oxford Website}, 
year = {2025}, 
url = {https://www.oxinst.com/products/ultra-low-temperature-platforms/} 
}

@misc{maybell, 
author = {{Maybell}}, 
title = {Maybell Website}, 
year = {2025}, 
url = {https://www.maybellquantum.com/} 
}

@misc{HeavyWaterReactor, 
author = {{HeavyWaterReactor}}, 
title = {Heavy Water Reactor Website}, 
year = {2025}, 
url = {https://cen.acs.org/materials/3-availability-increase-new-supply/99/web/2021/12} 
}

@misc{condor, 
author = {{condor}}, 
title = {condor Website}, 
year = {2024}, 
url = {https://www.livescience.com/technology/computing/ibm-scientists-built-massive-condor-1000-qubit-quantum-computer-chip-133-qubit-heron-system-two} 
}

@misc{yoder2025tourgrossmodularquantum,
      title={Tour de gross: A modular quantum computer based on bivariate bicycle codes}, 
      author={Theodore J. Yoder and Eddie Schoute and Patrick Rall and Emily Pritchett and Jay M. Gambetta and Andrew W. Cross and Malcolm Carroll and Michael E. Beverland},
      year={2025},
      eprint={2506.03094},
      archivePrefix={arXiv},
      primaryClass={quant-ph},
      url={https://arxiv.org/abs/2506.03094}, 
}

@misc{muller2025improvedbeliefpropagationsufficient,
      title={Improved belief propagation is sufficient for real-time decoding of quantum memory}, 
      author={Tristan Müller and Thomas Alexander and Michael E. Beverland and Markus Bühler and Blake R. Johnson and Thilo Maurer and Drew Vandeth},
      year={2025},
      eprint={2506.01779},
      archivePrefix={arXiv},
      primaryClass={quant-ph},
      url={https://arxiv.org/abs/2506.01779}, 
}

@article{Kern_2005,
   title={Quantum error correction of coherent errors by randomization},
   volume={32},
   ISSN={1434-6079},
   url={http://dx.doi.org/10.1140/epjd/e2004-00196-9},
   DOI={10.1140/epjd/e2004-00196-9},
   number={1},
   journal={The European Physical Journal D},
   publisher={Springer Science and Business Media LLC},
   author={Kern, O. and Alber, G. and Shepelyansky, D. L.},
   year={2005},
   month=jan, pages={153–156} }

@article{WarePFR2021,
  title = {Experimental Pauli-frame randomization on a superconducting qubit},
  author = {Ware, Matthew and Ribeill, Guilhem and Rist\`e, Diego and Ryan, Colm A. and Johnson, Blake and da Silva, Marcus P.},
  journal = {Phys. Rev. A},
  volume = {103},
  issue = {4},
  pages = {042604},
  numpages = {9},
  year = {2021},
  month = {Apr},
  publisher = {American Physical Society},
  doi = {10.1103/PhysRevA.103.042604},
  url = {https://link.aps.org/doi/10.1103/PhysRevA.103.042604}
}

@article{Knill_2005,
   title={Quantum computing with realistically noisy devices},
   volume={434},
   ISSN={1476-4687},
   url={http://dx.doi.org/10.1038/nature03350},
   DOI={10.1038/nature03350},
   number={7029},
   journal={Nature},
   publisher={Springer Science and Business Media LLC},
   author={Knill, E.},
   year={2005},
   month=mar, pages={39–44} }

@article{GST2009,
  doi = {10.22331/q-2021-10-05-557},
  url = {https://doi.org/10.22331/q-2021-10-05-557},
  title = {Gate {S}et {T}omography},
  author = {Nielsen, Erik and Gamble, John King and Rudinger, Kenneth and Scholten, Travis and Young, Kevin and Blume-Kohout, Robin},
  journal = {{Quantum}},
  issn = {2521-327X},
  publisher = {{Verein zur F{\"{o}}rderung des Open Access Publizierens in den Quantenwissenschaften}},
  volume = {5},
  pages = {557},
  month = oct,
  year = {2021}
}

@article{Ville_2022,
  title = {Leveraging randomized compiling for the quantum imaginary-time-evolution algorithm},
  author = {Ville, Jean-Loup and Morvan, Alexis and Hashim, Akel and Naik, Ravi K. and Lu, Marie and Mitchell, Bradley and Kreikebaum, John-Mark and O'Brien, Kevin P. and Wallman, Joel J. and Hincks, Ian and Emerson, Joseph and Smith, Ethan and Younis, Ed and Iancu, Costin and Santiago, David I. and Siddiqi, Irfan},
  journal = {Phys. Rev. Res.},
  volume = {4},
  issue = {3},
  pages = {033140},
  numpages = {10},
  year = {2022},
  month = {Aug},
  publisher = {American Physical Society},
  doi = {10.1103/PhysRevResearch.4.033140},
  url = {https://link.aps.org/doi/10.1103/PhysRevResearch.4.033140}
}

@article{Dugas_dihedral,
   title={Characterizing universal gate sets via dihedral benchmarking},
   volume={92},
   ISSN={1094-1622},
   url={http://dx.doi.org/10.1103/PhysRevA.92.060302},
   DOI={10.1103/physreva.92.060302},
   number={6},
   journal={Physical Review A},
   publisher={American Physical Society (APS)},
   author={Carignan-Dugas, Arnaud and Wallman, Joel J. and Emerson, Joseph},
   year={2015},
   month=dec }

@article{Knill_2008,
   title={Randomized benchmarking of quantum gates},
   volume={77},
   ISSN={1094-1622},
   url={http://dx.doi.org/10.1103/PhysRevA.77.012307},
   DOI={10.1103/physreva.77.012307},
   number={1},
   journal={Physical Review A},
   publisher={American Physical Society (APS)},
   author={Knill, E. and Leibfried, D. and Reichle, R. and Britton, J. and Blakestad, R. B. and Jost, J. D. and Langer, C. and Ozeri, R. and Seidelin, S. and Wineland, D. J.},
   year={2008},
   month=jan }

@article{Viola_1999,
   title={Dynamical Decoupling of Open Quantum Systems},
   volume={82},
   ISSN={1079-7114},
   url={http://dx.doi.org/10.1103/PhysRevLett.82.2417},
   DOI={10.1103/physrevlett.82.2417},
   number={12},
   journal={Physical Review Letters},
   publisher={American Physical Society (APS)},
   author={Viola, Lorenza and Knill, Emanuel and Lloyd, Seth},
   year={1999},
   month=mar, pages={2417–2421} }

@article{Ferracin_2024,
   title={Efficiently improving the performance of noisy quantum computers},
   volume={8},
   ISSN={2521-327X},
   url={http://dx.doi.org/10.22331/q-2024-07-15-1410},
   DOI={10.22331/q-2024-07-15-1410},
   journal={Quantum},
   publisher={Verein zur Forderung des Open Access Publizierens in den Quantenwissenschaften},
   author={Ferracin, Samuele and Hashim, Akel and Ville, Jean-Loup and Naik, Ravi and Carignan-Dugas, Arnaud and Qassim, Hammam and Morvan, Alexis and Santiago, David I. and Siddiqi, Irfan and Wallman, Joel J.},
   year={2024},
   month=jul, pages={1410} }

@article{BabbushEM,
  title = {Quantum error mitigation},
  author = {Cai, Zhenyu and Babbush, Ryan and Benjamin, Simon C. and Endo, Suguru and Huggins, William J. and Li, Ying and McClean, Jarrod R. and O'Brien, Thomas E.},
  journal = {Rev. Mod. Phys.},
  volume = {95},
  issue = {4},
  pages = {045005},
  numpages = {37},
  year = {2023},
  month = {Dec},
  publisher = {American Physical Society},
  doi = {10.1103/RevModPhys.95.045005},
  url = {https://link.aps.org/doi/10.1103/RevModPhys.95.045005}
}

@misc{Iyer2025,
      title={Enhancing Decoding Performance using Efficient Error Learning}, 
      author={Pavithran Iyer and Aditya Jain and Stephen D. Bartlett and Joseph Emerson},
      year={2025},
      eprint={2507.08536},
      archivePrefix={arXiv},
      primaryClass={quant-ph},
      url={https://arxiv.org/abs/2507.08536}, 
}

@misc{QCAP,
      title={Validating Quantum Computation via Circuit Benchmarking}, 
      author={Joseph Emerson and Joel Wallman and Arnaud-Carigna-Dugas},
      year={2025},
      eprint={Manuscript In Preparation},
      archivePrefix={arXiv},
      primaryClass={quant-ph}, 
}

@misc{Sannamoth2025,
      title={Easier randomizing gates provide more accurate fidelity estimation}, 
      author={Debankan Sannamoth and Kristine Boone and Arnaud Carigna-Dugas and Akel Hashim and Irfan Siddiqi and Karl Mayer  and Joseph Emerson},
      year={2025},
      eprint={Manuscript In Preparation},
      archivePrefix={arXiv},
      primaryClass={quant-ph}, 
}

@article{XEB,
   title={Characterizing quantum supremacy in near-term devices},
   volume={14},
   ISSN={1745-2481},
   url={http://dx.doi.org/10.1038/s41567-018-0124-x},
   DOI={10.1038/s41567-018-0124-x},
   number={6},
   journal={Nature Physics},
   publisher={Springer Science and Business Media LLC},
   author={Boixo, Sergio and Isakov, Sergei V. and Smelyanskiy, Vadim N. and Babbush, Ryan and Ding, Nan and Jiang, Zhang and Bremner, Michael J. and Martinis, John M. and Neven, Hartmut},
   year={2018},
   month=apr, pages={595–600} }

@article{CERforMCMb,
    author = "Zhang, Zhihan and Chen, Senrui and Liu, Yunchao and Jiang, Liang",
    title = "{Generalized Cycle Benchmarking Algorithm for Characterizing Midcircuit Measurements}",
    eprint = "2406.02669",
    archivePrefix = "arXiv",
    primaryClass = "quant-ph",
    doi = "10.1103/PRXQuantum.6.010310",
    journal = "PRX Quantum",
    volume = "6",
    number = "1",
    pages = "010310",
    year = "2025"
}

@article{CERforMCM,
  title = {Pauli Noise Learning for Mid-Circuit Measurements},
  author = {Hines, Jordan and Proctor, Timothy},
  journal = {Phys. Rev. Lett.},
  volume = {134},
  issue = {2},
  pages = {020602},
  numpages = {7},
  year = {2025},
  month = {Jan},
  publisher = {American Physical Society},
  doi = {10.1103/PhysRevLett.134.020602},
  url = {https://link.aps.org/doi/10.1103/PhysRevLett.134.020602}
}

@article{SimRB,
  title = {Three-Qubit Randomized Benchmarking},
  author = {McKay, David C. and Sheldon, Sarah and Smolin, John A. and Chow, Jerry M. and Gambetta, Jay M.},
  journal = {Phys. Rev. Lett.},
  volume = {122},
  issue = {20},
  pages = {200502},
  numpages = {6},
  year = {2019},
  month = {May},
  publisher = {American Physical Society},
  doi = {10.1103/PhysRevLett.122.200502},
  url = {https://link.aps.org/doi/10.1103/PhysRevLett.122.200502}
}

@misc{Gottesman2009introduction,
      title={An Introduction to Quantum Error Correction and Fault-Tolerant Quantum Computation}, 
      author={Daniel Gottesman},
      year={2009},
      eprint={0904.2557},
      archivePrefix={arXiv},
      primaryClass={quant-ph},
      url={https://arxiv.org/abs/0904.2557}, 
}

@article{GottesmanPhysRevA,
  title = {Theory of fault-tolerant quantum computation},
  author = {Gottesman, Daniel},
  journal = {Phys. Rev. A},
  volume = {57},
  issue = {1},
  pages = {127--137},
  numpages = {0},
  year = {1998},
  month = {Jan},
  publisher = {American Physical Society},
  doi = {10.1103/PhysRevA.57.127},
  url = {https://link.aps.org/doi/10.1103/PhysRevA.57.127}
}

@article{magesan2012characterizing,
  title={Characterizing quantum gates via randomized benchmarking},
  author={Magesan, Easwar and Gambetta, Jay M and Emerson, Joseph},
  journal={Physical Review A},
  volume={85},
  number={4},
  pages={042311},
  year={2012},
  publisher={APS},
  doi = {10.1103/PhysRevA.85.042311},
  url = {https://link.aps.org/doi/10.1103/PhysRevA.85.042311}
}

@article{magesan2012efficient,
  title={Efficient measurement of quantum gate error by interleaved randomized benchmarking},
  author={Magesan, Easwar and Gambetta, Jay M and Johnson, Blake R and Ryan, Colm A and Chow, Jerry M and Merkel, Seth T and Da Silva, Marcus P and Keefe, George A and Rothwell, Mary B and Ohki, Thomas A and others},
  journal={Physical Review Letters},
  volume={109},
  number={8},
  pages={080505},
  year={2012},
  publisher={APS}
}

@article{wallman2016noise,
  title={Noise tailoring for scalable quantum computation via randomized compiling},
  author={Wallman, Joel J and Emerson, Joseph},
  journal={Physical Review A},
  volume={94},
  number={5},
  pages={052325},
  year={2016},
  publisher={APS},
  doi = {10.1103/PhysRevA.94.052325},
  url = {https://link.aps.org/doi/10.1103/PhysRevA.94.052325}
}

@article{Emerson2007,
	Author = {Emerson, Joseph and Silva, Marcus and Moussa, Osama and Ryan, Colm and Laforest, Martin and Baugh, Jonathan and Cory, David G. and Laflamme, Raymond},
	Doi = {10.1126/science.1145699},
	Journal = {Science},
	Number = {5846},
	Pages = {1893-1896},
	Title = {Symmetrized Characterization of Noisy Quantum Processes},
	Volume = {317},
	Year = {2007}
}

@article{dankert2009exact,
  title={Exact and approximate unitary 2-designs and their application to fidelity estimation},
  author={Dankert, Christoph and Cleve, Richard and Emerson, Joseph and Livine, Etera},
  journal={Physical Review A},
  volume={80},
  number={1},
  pages={012304},
  year={2009},
  publisher={APS},
  doi = {10.1103/PhysRevA.80.012304},
  url = {https://link.aps.org/doi/10.1103/PhysRevA.80.012304}
}

@article{magesan2011scalable,
  title={Scalable and robust randomized benchmarking of quantum processes},
  author={Magesan, Easwar and Gambetta, Jay M and Emerson, Joseph},
  journal={Physical Review Letters},
  volume={106},
  number={18},
  pages={180504},
  year={2011},
  publisher={APS},
  doi = {10.1103/PhysRevLett.106.180504},
  url = {https://link.aps.org/doi/10.1103/PhysRevLett.106.180504}
}

@article{emerson2005scalable,
  title={Scalable noise estimation with random unitary operators},
  author={Emerson, Joseph and Alicki, Robert and {\.Z}yczkowski, Karol},
  journal={Journal of Optics B},
  volume={7},
  number={10},
  pages={S347},
  year={2005},
  publisher={IOP Publishing},
doi = {10.1088/1464-4266/7/10/021},
url = {https://dx.doi.org/10.1088/1464-4266/7/10/021}
}

@article{Hashim_2025,
   title={Practical Introduction to Benchmarking and Characterization of Quantum Computers},
   volume={6},
   ISSN={2691-3399},
   url={http://dx.doi.org/10.1103/PRXQuantum.6.030202},
   DOI={10.1103/prxquantum.6.030202},
   number={3},
   journal={PRX Quantum},
   publisher={American Physical Society (APS)},
   author={Hashim, Akel and Nguyen, Long B. and Goss, Noah and Marinelli, Brian and Naik, Ravi K. and Chistolini, Trevor and Hines, Jordan and Marceaux, J.P. and Kim, Yosep and Gokhale, Pranav and Tomesh, Teague and Chen, Senrui and Jiang, Liang and Ferracin, Samuele and Rudinger, Kenneth and Proctor, Timothy and Young, Kevin C. and Siddiqi, Irfan and Blume-Kohout, Robin},
   year={2025},
   month=aug }

@article{Helsen_2022,
   title={General Framework for Randomized Benchmarking},
   volume={3},
   ISSN={2691-3399},
   url={http://dx.doi.org/10.1103/PRXQuantum.3.020357},
   DOI={10.1103/prxquantum.3.020357},
   number={2},
   journal={PRX Quantum},
   publisher={American Physical Society (APS)},
   author={Helsen, J. and Roth, I. and Onorati, E. and Werner, A.H. and Eisert, J.},
   year={2022},
   month=jun }

@article{carignandugas2023error,
      title={The Error Reconstruction and Compiled Calibration of Quantum Computing Cycles}, 
      author={Arnaud Carignan-Dugas and Dar Dahlen and Ian Hincks and Egor Ospadov and Stefanie J. Beale and Samuele Ferracin and Joshua Skanes-Norman and Joseph Emerson and Joel J. Wallman},
      year={2023},
      journal={arXiv preprint arXiv:2303.17714},
      primaryClass={quant-ph},
      url={https://arxiv.org/abs/2303.17714}, 
}

@article{Carignan_Dugas_2019a,
   title={A polar decomposition for quantum channels (with applications to bounding error propagation in quantum circuits)},
   volume={3},
   ISSN={2521-327X},
   url={http://dx.doi.org/10.22331/q-2019-08-12-173},
   DOI={10.22331/q-2019-08-12-173},
   journal={Quantum},
   publisher={Verein zur Forderung des Open Access Publizierens in den Quantenwissenschaften},
   author={Carignan-Dugas, Arnaud and Alexander, Matthew and Emerson, Joseph},
   year={2019},
   month=aug, pages={173} }

@article{Carignan_Dugas_2019b,
   title={Bounding the average gate fidelity of composite channels using the unitarity},
   volume={21},
   ISSN={1367-2630},
   url={http://dx.doi.org/10.1088/1367-2630/ab1800},
   DOI={10.1088/1367-2630/ab1800},
   number={5},
   journal={New Journal of Physics},
   publisher={IOP Publishing},
   author={Carignan-Dugas, Arnaud and Wallman, Joel J and Emerson, Joseph},
   year={2019},
   month=may, pages={053016} }

@article{Foxen_2020,
   title={Demonstrating a Continuous Set of Two-qubit Gates for Near-term Quantum Algorithms},
   volume={125},
   ISSN={1079-7114},
   url={http://dx.doi.org/10.1103/PhysRevLett.125.120504},
   DOI={10.1103/physrevlett.125.120504},
   number={12},
   journal={Physical Review Letters},
   publisher={American Physical Society (APS)},
   author={Foxen, B. and Neill, C. and Dunsworth, A. and Roushan, P. and Chiaro, B. and Megrant, A. and Kelly, J. and Chen, Zijun and Satzinger, K. and Barends, R. and Arute, F. and Arya, K. and Babbush, R. and Bacon, D. and Bardin, J.C. and Boixo, S. and Buell, D. and Burkett, B. and Chen, Yu and Collins, R. and Farhi, E. and Fowler, A. and Gidney, C. and Giustina, M. and Graff, R. and Harrigan, M. and Huang, T. and Isakov, S.V. and Jeffrey, E. and Jiang, Z. and Kafri, D. and Kechedzhi, K. and Klimov, P. and Korotkov, A. and Kostritsa, F. and Landhuis, D. and Lucero, E. and McClean, J. and McEwen, M. and Mi, X. and Mohseni, M. and Mutus, J.Y. and Naaman, O. and Neeley, M. and Niu, M. and Petukhov, A. and Quintana, C. and Rubin, N. and Sank, D. and Smelyanskiy, V. and Vainsencher, A. and White, T.C. and Yao, Z. and Yeh, P. and Zalcman, A. and Neven, H. and Martinis, J.M.},
   year={2020},
   month=sep }

@article{Wallman_2014,
   title={Randomized benchmarking with confidence},
   volume={16},
   ISSN={1367-2630},
   url={http://dx.doi.org/10.1088/1367-2630/16/10/103032},
   DOI={10.1088/1367-2630/16/10/103032},
   number={10},
   journal={New Journal of Physics},
   publisher={IOP Publishing},
   author={Wallman, Joel J and Flammia, Steven T},
   year={2014},
   month=oct, pages={103032} }

@misc{Ceasura2022,
      title={Non-Exponential Behaviour in Logical Randomized Benchmarking}, 
      author={Athena Ceasura and Pavithran Iyer and Joel J. Wallman and Hakop Pashayan},
      year={2022},
      eprint={2212.05488},
      archivePrefix={arXiv},
      primaryClass={quant-ph},
      url={https://arxiv.org/abs/2212.05488}, 
}

@misc{EPLG,
      title={Benchmarking Quantum Processor Performance at Scale}, 
      author={David C. McKay and Ian Hincks and Emily J. Pritchett and Malcolm Carroll and Luke C. G. Govia and Seth T. Merkel},
      year={2023},
      eprint={2311.05933},
      archivePrefix={arXiv},
      primaryClass={quant-ph},
      url={https://arxiv.org/abs/2311.05933}, 
}

@article{Boone_2019,
   title={Randomized benchmarking under different gate sets},
   volume={99},
   ISSN={2469-9934},
   url={http://dx.doi.org/10.1103/PhysRevA.99.032329},
   DOI={10.1103/physreva.99.032329},
   number={3},
   journal={Physical Review A},
   publisher={American Physical Society (APS)},
   author={Boone, Kristine and Carignan-Dugas, Arnaud and Wallman, Joel J. and Emerson, Joseph},
   year={2019},
   month=mar }

@article{erhard2019characterizing,
  title={Characterizing large-scale quantum computers via cycle benchmarking},
  author={Erhard, Alexander and Wallman, Joel J and Postler, Lukas and Meth, Michael and Stricker, Roman and Martinez, Esteban A and Schindler, Philipp and Monz, Thomas and Emerson, Joseph and Blatt, Rainer},
  journal={Nature Communications},
  volume={10},
  number={1},
  pages={5347},
  year={2019},
  publisher={Nature Publishing Group UK London},
doi={10.1038/s41467-019-13068-7},
url={https://doi.org/10.1038/s41467-019-13068-7}
}

@article{chen2023learnability,
author={Chen, Senrui
and Liu, Yunchao
and Otten, Matthew
and Seif, Alireza
and Fefferman, Bill
and Jiang, Liang},
title= {The learnability of Pauli noise},
journal= {Nature Communications},
year={2023},
month={Jan},
day={04},
volume={14},
number={1},
pages={52},
issn={2041-1723},
doi={10.1038/s41467-022-35759-4},
url={https://doi.org/10.1038/s41467-022-35759-4}
}

@article{hockings2024scalable,
  title = {Scalable Noise Characterization of Syndrome-Extraction Circuits with Averaged Circuit Eigenvalue Sampling},
  author = {Hockings, Evan T. and Doherty, Andrew C. and Harper, Robin},
  journal = {PRX Quantum},
  volume = {6},
  issue = {1},
  pages = {010334},
  numpages = {34},
  year = {2025},
  month = {Feb},
  publisher = {American Physical Society},
  doi = {10.1103/PRXQuantum.6.010334},
  url = {https://link.aps.org/doi/10.1103/PRXQuantum.6.010334}
}

@article{iyer2025enhancing,
  title={Enhancing Decoding Performance using Efficient Error Learning},
  author={Iyer, Pavithran and Jain, Aditya and Bartlett, Stephen D and Emerson, Joseph},
  journal={arXiv preprint arXiv:2507.08536},
      url={https://arxiv.org/abs/2507.08536}, 
  year={2025}
}

@article{fazio2025,
      title={Characterizing physical and logical errors in a transversal CNOT via cycle error reconstruction}, 
      author={Nicholas Fazio and Robert Freund and Debankan Sannamoth and Alex Steiner and Christian D. Marciniak and Manuel Rispler and Robin Harper and Thomas Monz and Joseph Emerson and Stephen D. Bartlett},
      year={2025},
      eprint={2504.11980},
      archivePrefix={arXiv},
      journal={arXiv preprint arXiv:2504.11980},
      primaryClass={quant-ph},
      url={https://arxiv.org/abs/2504.11980}, 
}

@misc{HERC,
      title={Hardware-Efficient Randomized Compiling}, 
      author={Neelay Fruitwala and Akel Hashim and Abhi D. Rajagopala and Yilun Xu and Jordan Hines and Ravi K. Naik and Irfan Siddiqi and Katherine Klymko and Gang Huang and Kasra Nowrouzi},
      year={2024},
      eprint={2406.13967},
      archivePrefix={arXiv},
      primaryClass={quant-ph},
      url={https://arxiv.org/abs/2406.13967}, 
}

@misc{xu2023constantoverheadfaulttolerantquantumcomputation,
      title={Constant-Overhead Fault-Tolerant Quantum Computation with Reconfigurable Atom Arrays}, 
      author={Qian Xu and J. Pablo Bonilla Ataides and Christopher A. Pattison and Nithin Raveendran and Dolev Bluvstein and Jonathan Wurtz and Bane Vasic and Mikhail D. Lukin and Liang Jiang and Hengyun Zhou},
      year={2023},
      eprint={2308.08648},
      archivePrefix={arXiv},
      primaryClass={quant-ph},
      url={https://arxiv.org/abs/2308.08648}, 
}

@article{Roffe_PhysRevResearch.2.043423,
  title = {Decoding across the quantum low-density parity-check code landscape},
  author = {Roffe, Joschka and White, David R. and Burton, Simon and Campbell, Earl},
  journal = {Phys. Rev. Res.},
  volume = {2},
  issue = {4},
  pages = {043423},
  numpages = {13},
  year = {2020},
  month = {Dec},
  publisher = {American Physical Society},
  doi = {10.1103/PhysRevResearch.2.043423},
  url = {https://link.aps.org/doi/10.1103/PhysRevResearch.2.043423}
}

@article{Shaw_2025,
   title={Lowering Connectivity Requirements for Bivariate Bicycle Codes Using Morphing Circuits},
   volume={134},
   ISSN={1079-7114},
   url={http://dx.doi.org/10.1103/PhysRevLett.134.090602},
   DOI={10.1103/physrevlett.134.090602},
   number={9},
   journal={Physical Review Letters},
   publisher={American Physical Society (APS)},
   author={Shaw, Mackenzie H. and Terhal, Barbara M.},
   year={2025},
   month=mar }

@misc{he2025extractorsqldpcarchitecturesefficient,
      title={Extractors: QLDPC Architectures for Efficient Pauli-Based Computation}, 
      author={Zhiyang He and Alexander Cowtan and Dominic J. Williamson and Theodore J. Yoder},
      year={2025},
      eprint={2503.10390},
      archivePrefix={arXiv},
      primaryClass={quant-ph},
      url={https://arxiv.org/abs/2503.10390}, 
}

@article{van_den_Berg_2023,
   title={Probabilistic error cancellation with sparse Pauli–Lindblad models on noisy quantum processors},
   volume={19},
   ISSN={1745-2481},
   url={http://dx.doi.org/10.1038/s41567-023-02042-2},
   DOI={10.1038/s41567-023-02042-2},
   number={8},
   journal={Nature Physics},
   publisher={Springer Science and Business Media LLC},
   author={van den Berg, Ewout and Minev, Zlatko K. and Kandala, Abhinav and Temme, Kristan},
   year={2023},
   month=may, pages={1116–1121} }

@misc{chen2025efficientselfconsistentlearninggate,
      title={Efficient self-consistent learning of gate set Pauli noise}, 
      author={Senrui Chen and Zhihan Zhang and Liang Jiang and Steven T. Flammia},
      year={2025},
      eprint={2410.03906},
      archivePrefix={arXiv},
      primaryClass={quant-ph},
      url={https://arxiv.org/abs/2410.03906}, 
}

@misc{chen2025disambiguatingpaulinoisequantum,
      title={Disambiguating Pauli noise in quantum computers}, 
      author={Edward H. Chen and Senrui Chen and Laurin E. Fischer and Andrew Eddins and Luke C. G. Govia and Brad Mitchell and Andre He and Youngseok Kim and Liang Jiang and Alireza Seif},
      year={2025},
      eprint={2505.22629},
      archivePrefix={arXiv},
      primaryClass={quant-ph},
      url={https://arxiv.org/abs/2505.22629}, 
}

@article{Kim2023QuantumUtility,
  author    = {Youngseok Kim and Andrew Eddins and Sajant Anand and Ken Xuan Wei and Ewout van den Berg and Abhinav Kandala and Sami Rosenblatt and Hasan Nayfeh and Yantao Wu and Michael Zaletel and Kristan Temme},
  title     = {Evidence for the utility of quantum computing before fault tolerance},
  journal   = {Nature},
  volume    = {618},
  number    = {7966},
  pages     = {500--505},
  year      = {2023},
  doi       = {10.1038/s41586-023-06096-3},
  url       = {https://doi.org/10.1038/s41586-023-06096-3}
}

@inproceedings{ZettlesWJWA22,
  title = {26.2 Design Considerations for Superconducting Quantum Systems},
  author = {George Zettles and Scott Willenborg and Blake R. Johnson and Andrew Wack and Brian Allison},
  year = {2022},
  doi = {10.1109/ISSCC42614.2022.9731706},
  url = {https://doi.org/10.1109/ISSCC42614.2022.9731706},
  researchr = {https://researchr.org/publication/ZettlesWJWA22},
  cites = {0},
  citedby = {0},
  pages = {1-3},
  booktitle = {IEEE International Solid-State Circuits Conference, ISSCC 2022, San Francisco, CA, USA, February 20-26, 2022},
  publisher = {IEEE},
  isbn = {978-1-6654-2800-2},
}

@INPROCEEDINGS{Frank23ISSCC,
  author={Frank, David J. and Chakraborty, Sudipto and Tien, Kevin and Rosno, Pat and Yeck, Mark and Glick, Joseph A. and Robertazzi, Raphael and Richetta, Ray and Bulzacchelli, John F. and Ramirez, Daniel and Yilma, Dereje and Davies, Andrew and Joshi, Rajiv V. and Lekuch, Scott and Inoue, Ken and Underwood, Devin and Wisnieff, Dorothy and Baks, Chris and Timmerwilke, John and Song, Peilin and Johnson, Blake R. and Gaucher, Brian P. and Friedman, Daniel J.},
  booktitle={2023 IEEE Custom Integrated Circuits Conference (CICC)}, 
  title={Low power cryogenic RF ASICs for quantum computing}, 
  year={2023},
  volume={},
  number={},
  pages={1-8},
  doi={10.1109/CICC57935.2023.10121266}}

@INPROCEEDINGS{ParkIntel21,
  author={Park, Jong-Seok and Subramanian, Sushil and Lampert, Lester and Mladenov, Todor and Klotchkov, Ilya and Kurian, Dileep J. and Juarez-Hernandez, Esdras and Perez-Esparza, Brando and Kale, Sirisha Rani and Asma Beevi, K. T. and Premaratne, Shavindra and Watson, Thomas and Suzuki, Satoshi and Rahman, Mustafijur and Timbadiya, Jaykant B. and Soni, Saksham and Pellerano, Stefano},
  booktitle={2021 IEEE International Solid-State Circuits Conference (ISSCC)}, 
  title={13.1 A Fully Integrated Cryo-CMOS SoC for Qubit Control in Quantum Computers Capable of State Manipulation, Readout and High-Speed Gate Pulsing of Spin Qubits in Intel 22nm FFL FinFET Technology}, 
  year={2021},
  volume={64},
  number={},
  pages={208-210},
  doi={10.1109/ISSCC42613.2021.9365762}}

@ARTICLE{Chakraborty21,
  author={Chakraborty, Sudipto and Joshi, Rajiv V.},
  journal={IEEE Circuits and Systems Magazine}, 
  title={Cryogenic CMOS Design for Qubit Control: Present Status, Challenges, and Future Directions [Feature]}, 
  year={2024},
  volume={24},
  number={2},
  pages={34-46},
  doi={10.1109/MCAS.2024.3383808}}

@inproceedings{Kang2023CryogenicDRAG,
  author    = {Kiseo Kang and Donggyu Minn and Jaeho Lee and Ho-Jin Song and Moonjoo Lee and Jae-Yoon Sim},
  title     = {A Cryogenic Controller IC for Superconducting Qubits with DRAG Pulse Generation by Direct Synthesis without Using Memory},
  booktitle = {2023 IEEE International Solid-State Circuits Conference (ISSCC)},
  pages     = {33--35},
  year      = {2023},
  doi       = {10.1109/ISSCC42613.2023.10067245},
  publisher = {IEEE}
}

@article{Yoo2024CryogenicCMOS,
  author    = {J. Yoo and D. Minn and J. Lee and H.-J. Song and M. Lee and J.-Y. Sim},
  title     = {Cryogenic CMOS Design for Qubit Control: Present Status, Challenges, and Prospects},
  journal   = {IEEE Journal of Solid-State Circuits},
  volume    = {59},
  number    = {5},
  pages     = {1173--1185},
  year      = {2024},
  month     = {May},
  doi       = {10.1109/JSSC.2024.3352739},
  publisher = {IEEE}
}

@misc{maurer2025realtimedecodinggrosscode,
      title={Real-time decoding of the gross code memory with FPGAs}, 
      author={Thilo Maurer and Markus Bühler and Michael Kröner and Frank Haverkamp and Tristan Müller and Drew Vandeth and Blake R. Johnson},
      year={2025},
      eprint={2510.21600},
      archivePrefix={arXiv},
      primaryClass={quant-ph},
      url={https://arxiv.org/abs/2510.21600}, 
}

@article{Battistel2023RealTime,
  title={Real-Time Decoding for Fault-Tolerant Quantum Computing: Progress, Challenges and Outlook},
  author={Francesco Battistel and Christopher Chamberland and Kauser Johar and Ramon W. J. Overwater and Fabio Sebastiano and Luka Skoric and Yosuke Ueno and Muhammad Usman},
  journal={Quantum Science and Technology},
  volume={8},
  number={4},
  pages={045002},
  year={2023},
  doi={10.1088/2399-1984/aceba6},
  url={https://arxiv.org/abs/2303.00054}
}

@misc{eisert2025mindgapsfraughtroad,
      title={Mind the gaps: The fraught road to quantum advantage}, 
      author={Jens Eisert and John Preskill},
      year={2025},
      eprint={2510.19928},
      archivePrefix={arXiv},
      primaryClass={quant-ph},
      url={https://arxiv.org/abs/2510.19928}, 
}

@misc{zimborás2025mythsquantumcomputationfault,
      title={Myths around quantum computation before full fault tolerance: What no-go theorems rule out and what they don't}, 
      author={Zoltán Zimborás and Bálint Koczor and Zoë Holmes and Elsi-Mari Borrelli and András Gilyén and Hsin-Yuan Huang and Zhenyu Cai and Antonio Acín and Leandro Aolita and Leonardo Banchi and Fernando G. S. L. Brandão and Daniel Cavalcanti and Toby Cubitt and Sergey N. Filippov and Guillermo García-Pérez and John Goold and Orsolya Kálmán and Elica Kyoseva and Matteo A. C. Rossi and Boris Sokolov and Ivano Tavernelli and Sabrina Maniscalco},
      year={2025},
      eprint={2501.05694},
      archivePrefix={arXiv},
      primaryClass={quant-ph},
      url={https://arxiv.org/abs/2501.05694}, 
}

@misc{aharonov2025importanceerrormitigationquantum,
      title={On the Importance of Error Mitigation for Quantum Computation}, 
      author={Dorit Aharonov and Ori Alberton and Itai Arad and Yosi Atia and Eyal Bairey and Zvika Brakerski and Itsik Cohen and Omri Golan and Ilya Gurwich and Oded Kenneth and Eyal Leviatan and Netanel H. Lindner and Ron Aharon Melcer and Adiel Meyer and Gili Schul and Maor Shutman},
      year={2025},
      eprint={2503.17243},
      archivePrefix={arXiv},
      primaryClass={quant-ph},
      url={https://arxiv.org/abs/2503.17243}, 
}

@article{Dumitrescu2018CloudQC,
  title = {Cloud Quantum Computing of an Atomic Nucleus},
  author = {E. F. Dumitrescu and A. J. McCaskey and G. Hagen and G. R. Jansen and T. D. Morris and T. Papenbrock and R. C. Pooser and D. J. Dean and P. Lougovski},
  journal = {Physical Review Letters},
  volume = {120},
  number = {21},
  pages = {210501},
  year = {2018},
  doi = {10.1103/PhysRevLett.120.210501},
  publisher = {American Physical Society}
}

@misc{lanes2025frameworkquantumadvantage,
      title={A Framework for Quantum Advantage}, 
      author={Olivia Lanes and Mourad Beji and Antonio D. Corcoles and Constantin Dalyac and Jay M. Gambetta and Loic Henriet and Ali Javadi-Abhari and Abhinav Kandala and Antonio Mezzacapo and Christopher Porter and Sarah Sheldon and John Watrous and Christa Zoufal and Alexandre Dauphin and Borja Peropadre},
      year={2025},
      eprint={2506.20658},
      archivePrefix={arXiv},
      primaryClass={quant-ph},
      url={https://arxiv.org/abs/2506.20658}, 
}

@misc{yu2025quantumcentricalgorithmsamplebasedkrylov,
      title={Quantum-Centric Algorithm for Sample-Based Krylov Diagonalization}, 
      author={Jeffery Yu and Javier Robledo Moreno and Joseph T. Iosue and Luke Bertels and Daniel Claudino and Bryce Fuller and Peter Groszkowski and Travis S. Humble and Petar Jurcevic and William Kirby and Thomas A. Maier and Mario Motta and Bibek Pokharel and Alireza Seif and Amir Shehata and Kevin J. Sung and Minh C. Tran and Vinay Tripathi and Antonio Mezzacapo and Kunal Sharma},
      year={2025},
      eprint={2501.09702},
      archivePrefix={arXiv},
      primaryClass={quant-ph},
      url={https://arxiv.org/abs/2501.09702}, 
}

@article{RobledoMoreno2024Chemistry,
  title = {Chemistry beyond the scale of exact diagonalization on a quantum-centric supercomputer},
  author = {Robledo-Moreno, Javier and Motta, Mario and Haas, Holger and Javadi-Abhari, Ali and Jurcevic, Petar and Kirby, William and Martiel, Simon and Sharma, Kunal and Sharma, Sandeep and Shirakawa, Tomonori and Sitdikov, Iskandar and Sun, Rong-Yang and Sung, Kevin J. and Takita, Maika and Tran, Minh C. and Yunoki, Seiji and Mezzacapo, Antonio},
  journal = {Science Advances},
  volume = {10},
  number = {21},
  pages = {eadu9991},
  year = {2024},
  doi = {10.1126/sciadv.adu9991},
  url = {https://www.science.org/doi/10.1126/sciadv.adu9991}
}

@article{IEEE2023CryogenicElectronicsRoadmap,
  title = {Cryogenic Electronics and Quantum Information Processing},
  author = {IEEE International Roadmap for Devices and Systems},
  journal = {Institute of Electrical and Electronics Engineers},
  year = {2023},
  doi = {10.60627/042B-J892}
}

@ARTICLE{Krinner2019-ih,
  title    = "Engineering cryogenic setups for 100-qubit scale superconducting circuit systems",
  author   = "Krinner, S and Storz, S and Kurpiers, P and Magnard, P and Heinsoo, J and Keller, R and L{\"u}tolf, J and Eichler, C and Wallraff, A",
  journal  = "EPJ Quantum Technology",
  volume = 6,
  number = 1,
  month = may,
  year  = {2019},
  doi = {10.1140/epjqt/s40507-019-0072-0}
}

@misc{raicu2025cryogenicthermalmodelingmicrowave,
  title={Cryogenic Thermal Modeling of Microwave High Density Signaling}, 
  author={Naomi Raicu and Tom Hogan and Xian Wu and Mehrnoosh Vahidpour and David Snow and Matthew Hollister and Mark Field},
  year={2025},
  eprint={2502.01945},
  archivePrefix={arXiv},
  primaryClass={quant-ph},
  url={https://arxiv.org/abs/2502.01945}, 
  doi = {10.48550/arXiv.2502.01945}
}

@misc{manifold2025thermalcapacitymappingcryogenic,
  title={Thermal Capacity Mapping of Cryogenic Platforms for Quantum Computers}, 
  author={Scott A. Manifold and George B. Long and Jonathan J. Burnett},
  year={2025},
  eprint={2503.10775},
  archivePrefix={arXiv},
  primaryClass={quant-ph},
  url={https://arxiv.org/abs/2503.10775}, 
  doi = {10.48550/arXiv.2503.10775}
}

@article{Govia2023MidCircuitRB,
  title     = {A randomized benchmarking suite for mid-circuit measurements},
  author    = {L. C. G. Govia and P. Jurcevic and C. J. Wood and N. Kanazawa and S. T. Merkel and D. C. McKay},
  journal   = {New Journal of Physics},
  volume    = {25},
  number    = {12},
  pages     = {123016},
  year      = {2023},
  doi       = {10.1088/1367-2630/ad0e19},
  eprint    = {2207.04836},
  archivePrefix = {arXiv},
  primaryClass = {quant-ph}
}

@article{Shirizly2024DynamicRB,
  title     = {Randomized Benchmarking Protocol for Dynamic Circuits},
  author    = {Liran Shirizly and Luke C. G. Govia and David C. McKay},
  journal   = {arXiv preprint arXiv:2408.07677},
  year      = {2024},
  doi       = {10.48550/arXiv.2408.07677},
  archivePrefix = {arXiv},
  primaryClass = {quant-ph}
}

@article{B_umer_2024,
   title={Efficient Long-Range Entanglement Using Dynamic Circuits},
   volume={5},
   ISSN={2691-3399},
   url={http://dx.doi.org/10.1103/PRXQuantum.5.030339},
   DOI={10.1103/prxquantum.5.030339},
   number={3},
   journal={PRX Quantum},
   publisher={American Physical Society (APS)},
   author={Bäumer, Elisa and Tripathi, Vinay and Wang, Derek S. and Rall, Patrick and Chen, Edward H. and Majumder, Swarnadeep and Seif, Alireza and Minev, Zlatko K.},
   year={2024},
   month=aug }

@article{B_umer_2025,
   title={Measurement-based long-range entangling gates in constant depth},
   volume={7},
   ISSN={2643-1564},
   url={http://dx.doi.org/10.1103/PhysRevResearch.7.023120},
   DOI={10.1103/physrevresearch.7.023120},
   number={2},
   journal={Physical Review Research},
   publisher={American Physical Society (APS)},
   author={Bäumer, Elisa and Woerner, Stefan},
   year={2025},
   month=may }

@article{Carrera_Vazquez_2024,
   title={Combining quantum processors with real-time classical communication},
   volume={636},
   ISSN={1476-4687},
   url={http://dx.doi.org/10.1038/s41586-024-08178-2},
   DOI={10.1038/s41586-024-08178-2},
   number={8041},
   journal={Nature},
   publisher={Springer Science and Business Media LLC},
   author={Carrera Vazquez, Almudena and Tornow, Caroline and Ristè, Diego and Woerner, Stefan and Takita, Maika and Egger, Daniel J.},
   year={2024},
   month=nov, pages={75–79} }

@article{Kang_2025,
   title={Teleporting two-qubit entanglement across 19 qubits on a superconducting quantum computer},
   volume={23},
   ISSN={2331-7019},
   url={http://dx.doi.org/10.1103/PhysRevApplied.23.014057},
   DOI={10.1103/physrevapplied.23.014057},
   number={1},
   journal={Physical Review Applied},
   publisher={American Physical Society (APS)},
   author={Kang, Haiyue and Kam, John F. and Mooney, Gary J. and Hollenberg, Lloyd C.L.},
   year={2025},
   month=jan }

@article{hothem2024measuring,
  title={Measuring error rates of mid-circuit measurements},
  author={Hothem, Daniel and Hines, Jordan and Baldwin, Charles and Gresh, Dan and Blume-Kohout, Robin and Proctor, Timothy},
  journal={arXiv preprint arXiv:2410.16706},
  year={2024},
  doi={10.48550/arXiv.2410.16706}
}

@article{hines2024fully,
  title={Fully scalable randomized benchmarking without motion reversal},
  author={Hines, Jordan and Hothem, Daniel and Blume-Kohout, Robin and Whaley, Birgitta and Proctor, Timothy},
  journal={arXiv preprint arXiv:2309.05147},
  year={2024},
  doi={10.48550/arXiv.2309.05147}
}

@article{merkel2025clifford,
  title={When Clifford benchmarks are sufficient; estimating application performance with scalable proxy circuits},
  author={Merkel, Seth and Proctor, Timothy and Ferracin, Samuele and Hines, Jordan and Barron, Samantha and Govia, Luke C. G. and McKay, David},
  journal={arXiv preprint arXiv:2503.05943},
  year={2025},
  doi={10.48550/arXiv.2503.05943}
}

@misc{kanazawa2025observabilityarchitecturequantumcentricsupercomputing,
      title={Observability Architecture for Quantum-Centric Supercomputing Workflows}, 
      author={Naoki Kanazawa and Yuto Morohoshi and Hitomi Takahashi and Yukio Kawashima and Hiroshi Horii and Kengo Nakajima},
      year={2025},
      eprint={2512.05484},
      archivePrefix={arXiv},
      primaryClass={quant-ph},
      url={https://arxiv.org/abs/2512.05484}, 
}

@misc{javadiabhari2024quantumcomputingqiskit,
      title={Quantum computing with Qiskit}, 
      author={Ali Javadi-Abhari and Matthew Treinish and Kevin Krsulich and Christopher J. Wood and Jake Lishman and Julien Gacon and Simon Martiel and Paul D. Nation and Lev S. Bishop and Andrew W. Cross and Blake R. Johnson and Jay M. Gambetta},
      year={2024},
      eprint={2405.08810},
      archivePrefix={arXiv},
      primaryClass={quant-ph},
      url={https://arxiv.org/abs/2405.08810}, 
}

\end{document}